\documentclass{achemso}
\usepackage[version=3]{mhchem}
\usepackage[utf8]{inputenc}

\usepackage{titlecaps}
\Addlcwords{the of into}
\usepackage{chemformula} 
\usepackage[T1]{fontenc} 
\usepackage{enumitem}
\usepackage{graphics}
\usepackage{bm}
\usepackage[normalem]{ulem}
\usepackage[symbol]{footmisc}
\usepackage{multirow}
\usepackage[normalem]{ulem}
\SectionNumbersOn

\usepackage{makecell}

\usepackage{soul}
\usepackage{xr}
\makeatletter
\newcommand*{\addFileDependency}[1]{
  \typeout{(#1)}
  \@addtofilelist{#1}
  \IfFileExists{#1}{}{\typeout{No file #1.}}
 }
\makeatother

\newcommand*{\myexternaldocument}[1]{%
     \externaldocument{#1}%
     \addFileDependency{#1.tex}%
     \addFileDependency{#1.aux}%
 }
\myexternaldocument{si}

\author{JingChun Wang, Meenu Upadhyay, Eric D. Boittier, Kham Lek
  Chaton, Valerii Andreichev, Mike Devereux, Shimoni Patel, Sena
  Aydin} \affiliation{Department of Chemistry, University of Basel,
  Klingelbergstrasse 80, CH-4056 Basel, Switzerland} \author{Kai
  T\"{o}pfer} \affiliation{Department of Chemistry, University of
  Basel, Klingelbergstrasse 80, CH-4056 Basel, Switzerland}
\author{Markus Meuwly}\email{m.meuwly@unibas.ch}
\affiliation{Department of Chemistry, University of Basel,
  Klingelbergstrasse 80, CH-4056 Basel, Switzerland}

\title[]{Cluster Models for Next-Generation, Machine-Learning-Based
  Energy Functions for Molecular Simulations}

\begin{document}

\begin{abstract}
Energy functions for pure and heterogenous systems are one of the
backbones for molecular simulation of condensed phase systems. With
the advent of machine learned potential energy surfaces (ML-PESs) a
new era has started. Statistical models allow the representation of
reference data from electronic structure calculations for chemical
systems of almost arbitrary complexity at unprecedented detail and
accuracy. Here, kernel- and neural network-based approaches for
intramolecular degrees of freedom are combined with distributed charge
models for long range electrostatics to describe the interaction
energies of condensed phase systems. The main focus is on illustrative
examples ranging from pure liquids (dichloromethane, water) to
chemically and structurally heterogeneous systems (eutectic liquids,
CO on amorphous solid water), reactions (Menshutkin), and spectroscopy
(triatomic probes for protein dynamics). For all examples, small to
medium-sized clusters are used to represent and improve the total
interaction energy compared with reference quantum chemical
calculations. Although remarkable accuracy can be achieved for some
systems (chemical accuracy for dichloromethane and water), it is clear
that more realistic models are required for van der Waals
contributions and improved water models need to be used for more
quantitative simulations of heterogeneous chemical and biological
systems.
\end{abstract}

\section{Introduction}
Cluster systems consisting of a finite number of atomic and/or
molecular building blocks constitute a state of matter between
isolated gas phase units and the bulk. In the past, one of the main
driving forces to generate, investigate, and characterize such systems
was the realization that clusters of increasing sizes may provide an
understanding for how condensed phase properties emerge across various
length scales.\cite{keesee:1986} Furthermore, following the properties
of clusters as they increase in size may provide information about the
phenomenon of nucleation.\cite{keesee:1988} Historically, one of the
earliest examples that was investigated are metal
clusters.\cite{cotton:1964}\\

\noindent
Clusters as finite-sized aggregates of identical atoms/molecules or
mixed clusters have been investigated with great success.  For
example, spectroscopic and computational work on protonated water
clusters of increasing size has provided fundamental insights into the
relationship between structure and spectroscopy of such
systems.\cite{shin:2004,headrick:2005} Likewise, the structure,
energetics and thermodynamics of atomic clusters interacting through
energy functions exhibiting different strengths and range was
investigated.\cite{wales:1996} From an analysis of the distribution of
low-lying minima, some unusual thermodynamic properties of finite
systems were determined.\cite{hill:1963,wales:2018} Disconnectivity
graphs\cite{becker:1997} provided a compact rendering of the PES, the
relationship between minima and the connectivity through transition
states depending on the strength of the interatomic
interactions.\cite{wales:2018} More specifically, for pure and mixed
rare gas clusters the collision induced absorption spectra were
determined from classical molecular dynamics (MD) simulations and path
integral MD simulations were used to characterize the structures of
such clusters.\cite{MM.rg:2002,MM.rg:2010} The results from such
computations agreed favourably with experiments on their spectroscopy
and structure. All these investigations pointed towards a pronounced
dependence of the measured properties on the structure and size of the
systems.\\

\noindent
Clustering can also occur in solution. For example, the phenomenon of
microheterogeneity occurs in water/alcohol mixtures and leads to local
aggregation and nonuniform distribution of one type of species
although the mixture on larger than molecular length scales appears
more uniform.\cite{pozar:2016} Related aggregation and clustering
phenomena are also relevant to and at play in separation methods such
as high performance liquid chromatography
(HPLC).\cite{dorsey:1989,MM.chroma:2008,MM.chroma:2017,MM.chroma:2018}\\

\noindent
Parametrization of empirical energy functions (or force fields)
typically hinges on a combination of fitting to experimental reference
data and information obtained from electronic structure
calculations.\cite{cgenff:2010,lemkul:2022,MM.ff:2024,MM.jcc:2025} In
addition to monomer properties, monomer--water complexes are included
for parametrizing the CHARMM General Force Field(CGenFF) force field
for realistically describing H-bonding interactions.\cite{cgenff:2010}
In the case of CGenFF the water model used was the TIP3P
model.\cite{tip3p} Typical experimental observables are the gas phase
vibrational spectrum (infrared and/or Raman), the pure liquid density,
heat of vaporization, or diffusion coefficients in one-component
liquids. Information that needs to be obtained from {\it ab initio}
calculations are the partial charges on each of the atoms for which
different approaches exist.\cite{poater:2020} Estimating partial
charges from experimental X-ray crystallography is in principle
possible but requires highest-resolution structures.\cite{jelsch:2000}
In addition, {\it ab initio} calculations of clusters include local
information on how interaction energies depend on atomic-scale details
that are missing from (macroscopic) experimental data.\\

\noindent
It is of interest to note that J. E. (Lennard-)Jones, after whom
"Lennard-Jones clusters" are named and who researched the distance
dependence of the intermolecular interactions between weakly
interacting particles, wrote\cite{jones:1924} already in 1924 (italics
added): "Until our knowledge of the disposition and motion of the
electrons in atoms and molecules is more complete, we cannot hope to
make a direct calculation of the nature of the {\it forces} called
into play during an encounter between molecules in a gas. It is true
that [..] Debye [..]investigated the nature of the field in the
neighbourhood of a hydrogen atom [..] and has shown how the pulsating
field gives rise on the whole to a force of repulsion, as well as one
of attraction on a unit negative charge. But it is difficult to see
how this work can be extended to more complex systems.[..] One such
method is to {\it assume a definite law of force}, and then by the
methods of the kinetic theory to deduce the appropriate law of
dependence of the viscosity of a gas on temperature." Obviously,
Lennard-Jones thought of "force" instead of "energy", although the
expressions that were parametrized described how the total energy
changes with geometry.\\

\noindent
In the field of machine learning potential energy surfaces encoding
the chemical environment into the model plays a central role. Defining
such an environment is usually done by choosing a cutoff radius which
also leads to clusters of atoms surrounding a reference atom in order
to generate a representation - or descriptor - suitable for fitting a
machine learned PES.\cite{bartok:2013} One such descriptor is the
Smooth Overlap of Atomic Positions (SOAP) that quantifies the
similarity between any two neighbourhood environments, and its
performance was tested in particular on small silicon
clusters. Another possibility is to encode the environment through
features which are trained by minimizing a suitable loss
function. This is the approach followed in
PhysNet.\cite{MM.physnet:2019}\\

\noindent
The present work considers pure and mixed clusters as a test-bed to
develop accurate energy functions for condensed-phase simulations. The
approach taken here uses a combination of machine learning-based
techniques, combined with empirical expressions for the total energy
of the system. Such an approach provides flexibility, accuracy and
computational speed which are important for large(r)-scale
simulations. First, the methods used are described, followed by new
results on a range of paradigmatic systems, including pure substances,
adsorbates on water, mixed and electrostatically dominated mixtures,
reactive systems, and spectroscopic probes for condensed phase
systems.\\

\section{Methods}
This chapter briefly describes the energy functions employed in the
present work. More technical details are provided in each of the
results subsections together with specifics of the MD simulations
carried out, if applicable. All simulations were run with the CHARMM
program with provisions to use machine learned energy
functions.\cite{charmm:2009,MM.rkhs:2017,MM.physnet:2023,MM.charmm:2024}
If not otherwise mentioned, the empirical energy function used in the
present work is CGenFF\cite{cgenff:2010} together with the TIP3P water
model\cite{tip3p} for consistency.\\

\subsection{Machine-Learned Energy Functions}
The machine learning-based techniques used in the present work include
kernel- and neural network-based approaches.\\

\noindent
One powerful method to construct accurate PESs uses reproducing kernel
Hilbert spaces (RKHSs)\cite{rabitz:1999} for which dedicated code has
been made available.\cite{MM.rkhs:2017} The theory of reproducing
kernel Hilbert spaces asserts that for $N$ training values $f_i =
f(\mathbf{x}_i)$ of a function $f(\mathbf{x})$ at locations
$\mathbf{x}_i$, $f(\mathbf{x})$ at arbitrary position $\mathbf{x}$ can
always be approximated as a linear combination of kernel products
\cite{scholkopf2001generalized}
\begin{equation}
\widetilde{f}(\mathbf{x}) = \sum_{i = 1}^{N} c_i K(\mathbf{x},\mathbf{x}_i)
\label{eq:RKHS_function}
\end{equation}
Here, the $c_i$ are coefficients and $K(\mathbf{x},\mathbf{x'})$ is
the reproducing kernel of the RKHS. The coefficients $c_i$ satisfy the
linear relation
 \begin{equation}
 f_j = \sum_{i = 1}^{N} c_i K_{ij}
 \label{eq:RKHS_coefficient_relation}
 \end{equation}
where $i$ and $j$ are both elements of the training set and the
symmetric, positive-definite kernel matrix $K_{ij} =
K(\mathbf{x}_i,\mathbf{x}_j)$ and $c_{i}$ can therefore be calculated
from the known training values $f_i$ by solving
Eq.~\ref{eq:RKHS_coefficient_relation} for the unknowns $c_i$ using,
e.g.\ Cholesky decomposition.\cite{golub2012matrix} With the
coefficients $c_i$ determined, the function value at an arbitrary
position $\mathbf{x}$ can be calculated using
Eq.~\ref{eq:RKHS_function}. Derivatives of $\widetilde{f}(\mathbf{x})$
of any order can be calculated analytically by replacing the kernel
function $K(\mathbf{x},\mathbf{x'})$ in Eq.~\ref{eq:RKHS_function}
with its corresponding derivative.\\

\noindent
The explicit form of the multi-dimensional kernel function
$K(\mathbf{x},\mathbf{x'})$ also depends on the problem to be
solved. It is possible to construct $D$-dimensional kernels as tensor
products of one-dimensional kernels $k(x,x')$
\begin{equation}
K(\mathbf{x},\mathbf{x'}) = \prod_{d=1}^{D} k^{(d)}(x^{(d)},x'^{(d)})
\label{eq:multidimensional_kernel}
\end{equation}
For the kernel functions $k(x,x')$ explicit physical knowledge can be
encoded, for example the correct asymptotic decay for the long range
interactions.\cite{soldan:2000} Explicit radial kernels include the
reciprocal power decay kernel\cite{ho1996general}
\begin{equation}
k_{n,m}(x,x') = n^2
x_{>}^{-(m+1)}\mathrm{B}(m+1,n)_2\mathrm{F}_1\left(-n+1,m+1;n+m+1;\dfrac{
  x_{<}}{x_{>}}\right)
\label{eq:reciprocal_power_kernel}
\end{equation}
where $x_{>}$ and $x_{<}$ are the larger and smaller of $x$ and $x'$,
the integer $n$ determines the smoothness and $m$ controls the
long-range decay (e.g.\ $m=5$ for dispersion), $\mathrm{B}(a,b)$ is
the beta function and $_2\mathrm{F}_1(a,b;c;d)$ is the Gauss
hypergeometric function.\\

\noindent
The NN-based ML-PESs were trained using the PhysNet
architecture.\cite{MM.physnet:2019} The loss function to be optimized
included energies ($E^{\text{ref}}$), forces
($F^{\text{ref}}_{i,\alpha}$), the total charge ($Q^{\text{ref}}$),
and dipole moments ($p^{\text{ref}}_{\alpha}$) for $N$ training
structures
\begin{equation}
  \label{eq:loss}
\begin{aligned}
\mathcal{L} &= w_E \left| E - E^{\text{ref}} \right| + \frac{w_F}{3N}
\sum_{i=1}^{N} \sum_{\alpha=1}^{3} \left| -\frac{\partial E}{\partial
  r_{i,\alpha}} - F^{\text{ref}}_{i,\alpha} \right| \\ &+ w_Q \left|
\sum_{i=1}^{N} q_i - Q^{\text{ref}} \right| + \frac{w_p}{3}
\sum_{\alpha=1}^{3} \left| \sum_{i=1}^{N} q_i r_{i,\alpha} -
p^{\text{ref}}_{\alpha} \right| + \mathcal{L}_{\text{nh}}.
\end{aligned}
\end{equation}\\
and was minimized using the Adam
optimizer,\cite{kingma:2014,reddi:2019} and $\alpha$ refers to the
three Cartesian components of the vectorial quantities,
respectively. The
hyperparameters\cite{MM.physnet:2019,MM.physnet:2023} $w_i$ $i \in \{
E, F, Q, p \}$ differentially weigh the contributions to the loss
function and were $w_E = 1$, $w_F \sim 52.92$, $w_Q \sim 14.39$ and
$w_p \sim 27.21$, respectively, and the term $\mathcal{L}_{\text{nh}}$
is a ``nonhierarchical penalty'' that regularizes the loss
function.\cite{MM.physnet:2019}\\

\noindent
PhysNet belongs to the family of message-passing NNs (MPNNs) which
falls within the broader category of graph neural
networks.\cite{scarselli:2009} As with all MPNNs, PhysNet contains an
input layer, several hidden layers ("modules") and one output
layer. Each module in PhysNet consists of an interaction block and
several residual blocks to facilitate training as the depth of the NN
increases. In the present work, 5 hidden layers were used as in
previous work.\cite{MM.tl:2021,MM.criegee:2021} Based on nuclear
charges (the "chemistry") \textbf{$Z$} and positions $\mathbf{R}$ of
all atoms of a molecule, the feature vectors describing each atom in a
local chemical environment are iteratively refined, given total
energies and forces for a number of reference
structures.\cite{MM.physnet:2019,MM.physnet:2023,MM.review:2023,MM.charmm:2024}
The components of the feature vectors, which have length 128
throughout this work, were randomly initialized between $-\sqrt{3}$
and $\sqrt{3}$. As the messages propagate through the NN, the atomic
feature vectors are refined to minimize the total loss function
(Eq. \ref{eq:loss}).\cite{MM.physnet:2019}\\

\subsection{Anisotropic Electrostatics}
Models for electrostatics that go beyond atom-centered point charges
include atom-centered multipoles (MTP), minimally distributed charge
models (MDCM),\cite{MM.mdcm:2017} and conformationally dependent MDCM
where the dependence is either described by explicit parametrized
functions (fMDCM)\cite{MM.fmdcm:2022} or through 1-dimensional kernel
functions (kMDCM) that act on internal
coordinates.\cite{MM.kmdcm:2024} For fMDCM models, a single internal
degree of freedom, often a valence angle, parametrizes the position of
a number of distributed charges in the local (molecular) frame, which
is achieved through constrained least-squares fitting to the ESP for
several distorted structures. The functions chosen to parameterize
these `flexible' charges are arbitrary but a polynomial expansion up
to the third power is generally sufficient. This addition of an
internal polarization contribution complements external polarization
models such as the
Drude-treatment\cite{MacKerell:2003,MacKerell:2016,Brooks:2017,MacKerell:2025}
but specifically tuned to reproduce the electrostatic potential in the
van der Waals region of the molecule. The kMDCM model, used for water,
was an extension\cite{MM.kmdcm:2024} which generates optimized
non-equilibrium charge models using a Gaussian kernel-based
representation to describe anisotropic electrostatics which adapt
smoothly with molecular geometry. Further details on the fitting
procedure can be found in Ref. \citenum{MM.kmdcm:2024}.\\

\section{Results}
This section presents results for a range of systems with a focus on
improvements of the total energy function. For a few cases, the
sensitivity of computed observables on different models is explored to
highlight changes in the performance depending on the
parametrization.\\

\subsection{Dichloromethane}
Dichloromethane (DCM, CH$_2$Cl$_2$) is a widely used and extensively
studied solvent.\cite{balint:2007} Its intermolecular interactions are
dominated by dispersion, short-range repulsion and electrostatic
contributions from
H-bonding.\cite{Almasy:2019,allen:2013}. Traditional empirical energy
functions such as CHARMM,\cite{cgenff:2010} Amber,\cite{amber}
Gromos,\cite{gromos} and OPLS\cite{opls}, which rely on atom-centered
point charges and pairwise-additive non-bonded terms are limited to
representing isotropic charge distributions. Such models inadequately
describe anisotropic features such as $\sigma$-holes characteristic of
halogen atoms, including chlorine.\cite{clark:2007,MM.mtp:2016} Over
time, numerous models have been developed to more realistically
describe the non-bonded interactions in DCM, beginning with three- and
five-site models for electrostatics\cite{BALINT:1982} and five-site
Lennard–Jones representations.\cite{Evans:1982} Further refinements
have attempted to capture anisotropy in the bulk phase, including the
use of atomic quadrupolar moments,\cite{Torii:2005} effective pair
potentials,\cite{Bohm:1985} polarizable pair
potentials,\cite{Dang:1999} and re-parameterized van der Waals terms
to better describe solvation.\cite{Pollice:2017,savoy:2025} Building
on these developments, the present work explores the use of machine
learning models as a cost-efficient alternative to explicitly complex
functional terms. In particular, machine-learned dimer potentials can
provide accurate short-range interaction energies, with an appropriate
cutoff allowing a smooth transition to a more accurate electrostatic
model or CGenFF charges at medium and long range. In this way,
empirical force fields can be systematically reparametrized and
extended to yield a more transferable and physically grounded
description of DCM. The use of an additive dimer potential with no
explicit many-body correction for bulk simulations is justified for
DCM due to relatively weak H-bonding\cite{allen:2013} and polarization
contributions that limit many-body effects\cite{MM.ff:2024}.\\

\noindent
To generate clusters for parametrization, a $32^3$ \AA\/$^3$ cubic box
was generated using PACKMOL\cite{martinez.pm:2009}. Using
CHARMM\cite{MM.charmm:2024} and the CGenFF\cite{cgenff:2010} energy
function, the system was relaxed through 2000 steps of Steepest
Descent (SD) algorithm, followed by heating and equilibration
simulations of 20 ps and 50 ps, respectively, with a timestep of
$\Delta t = 1$ fs, in the $NVT$ and $NPT$ ensembles. A Nos\'e-Hoover
thermostat was used to maintain the temperature at 300 K and the total
pressure was maintained at 1 atm using the Langevin Piston
barostat\cite{feller:1995}. Long range interactions were treated using
particle mesh Ewald with a cutoff of 14 \AA\/ and the
Lennard-Jones interactions were switched between 10 \AA\/ and 12
\AA\/. From a 1 ns production simulations in the $NPT$ ensemble using
the Leapfrog Verlet integrator with a timestep of $\Delta t = 0.2$ fs
every $100^{th}$ snapshot was saved. A total of 200 distinct clusters
containing 20 DCM molecules were extracted by randomly choosing a DCM
molecule and the 19 closest neighbors. For each cluster, the total
energy of DCM$_{20}$ and all 190 dimer energies DCM--DCM were
determined at the DLPNO-MP2/cc-pVTZ
level\cite{pinski:dlpno:2015,neugebauer:dlpno_2023} using
ORCA\cite{ORCA,ORCA5}.\\

\begin{figure}[h!]
\centering
\includegraphics[width=0.99\textwidth]{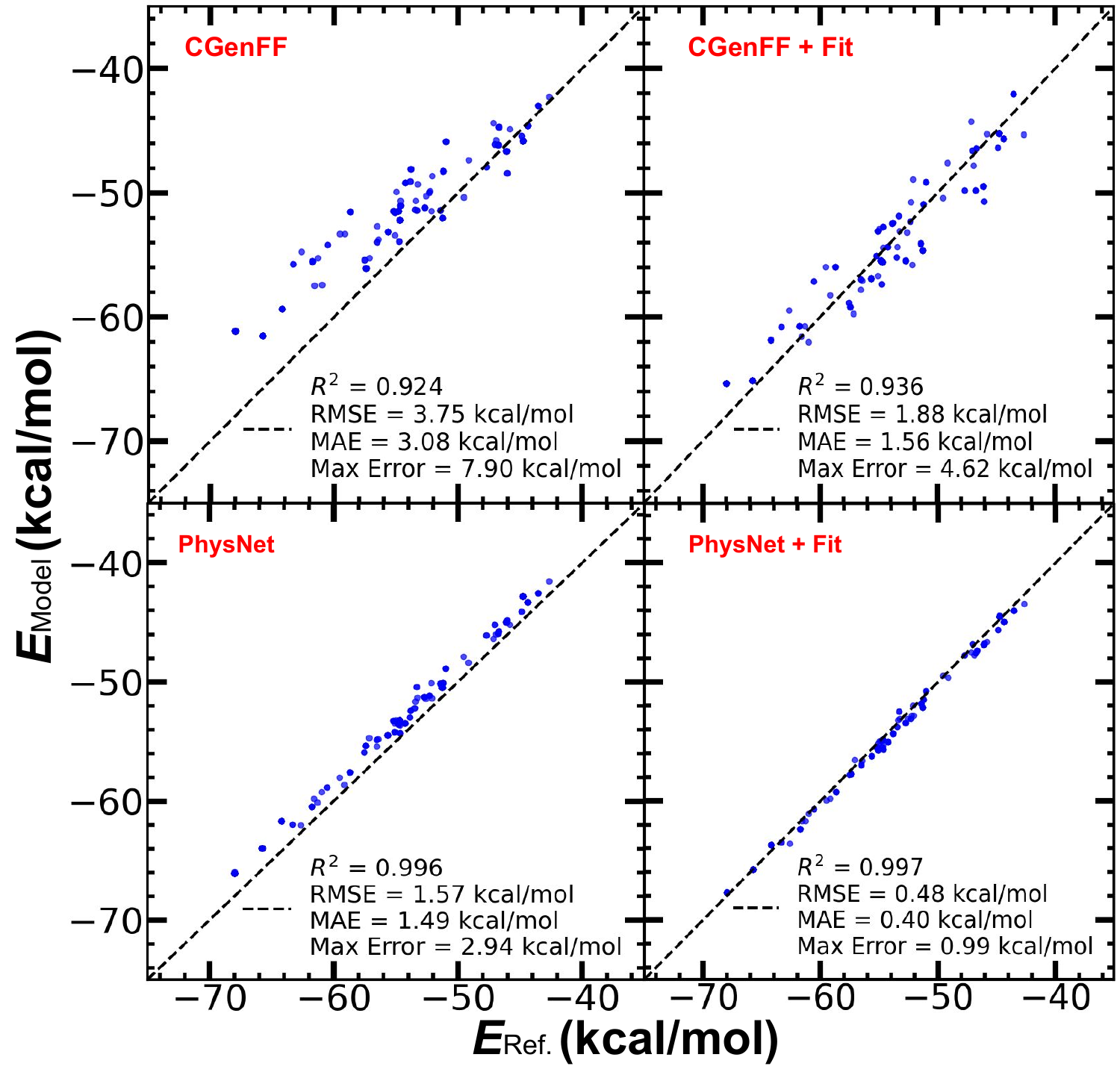}
\caption{Correlation Plot of Model (cluster) energies vs reference
  energies (Cluster). The cluster $(n = 20)$ geometries were extracted
  from an MD simulation run on CGenFF. The left panels have been
  marked by the respective energy functions used to get the energies
  of the clusters, and the Lennard-Jones potential was fitted for the
  corresponding right panels. The Model cluster energy is the
  non-bonded energy calculated by CHARMM, see panel "CGenFF". The
  PhysNet cluster energy is the sum of the dimer-pair energy in the
  corresponding cluster. The sum of dimer pairs is being correlated
  with the ORCA Formation Energy. The reference energies ($E_{\rm
    Ref.}$) were obtained from ORCA calculations done at the
  DLPNO-MP2-cc-pVTZ level of theory. The ORCA Formation energy in the
  X-axis was obtained following Eq. \ref{eq1}.}
\label{fig:dcm-nano} 
\end{figure}

\noindent
First, the total interaction energy from the empirical energy function
(CGenFF) is considered and improved by readjusting the
LJ-parameters. For this, $E^{\rm inter}$ is defined as the total
interaction energy of a given cluster from which the sum of monomer
energies were subtracted. The interaction energy $E^{\rm inter}_{j}$ of
cluster $j$ with $j \in[1,200]$ was obtained from the total energy
$E^{\rm total}_{j}$ from which the sum of the 20 monomer energies
$\sum_{i=1}^{20} E^{{\rm monomer}}_{i,j}$ were subtracted, see
Eq. \ref{eq1}. Hence, the interaction energy
\begin{equation}
E_{j}^{\rm inter}= E^{{\rm cluster}}_{j} - \sum_{i=1}^{20} E^{{\rm monomer}}_{i,j}
\label{eq1}
\end{equation}
from the electronic structure calculations can be directly compared
with the external/non-bonded energy contribution for all 200 clusters,
see top panels in Figure \ref{fig:dcm-nano}.\\

\noindent
Machine learned models like
PhysNet\cite{MM.physnet:2019,MM.physnet:2023} are well-suited for
describing close range intermolecular interactions but require a large
amount of training data when the distance between monomers
increases. Hence, combining an accurate close-range representation for
dimers using a NN-PES with a more empirical long-range representation
based on electrostatics and LJ-contributions is a potentially
data-efficient and accurate route. For constructing the dimer-PES, a
20-monomer DCM cluster $j$ contains 190 distinct dimer pairs. The
total formation energy for cluster $j$ can therefore be written as the
sum of these 190 dimer contributions, plus a residual term accounting
for many-body contributions
\begin{equation}
    E_{j}^{\rm inter} = \sum_{i=N}^{190} E_{i,j}^{\rm dimer, inter} +
    E_{j}^{\rm residual}
    \label{eq2}
\end{equation}
where 
\begin{equation}
     E_{i}^{\rm dimer, inter} = E^{\mathrm{dimer}}_{i} - \big(
     E^{\mathrm{monomer}}_{a,i} + E^{\mathrm{monomer}}_{b,i} \big)
     \label{eq3}
\end{equation}
Here, the residual energy contribution was not determined explicitly
and hence $E_{j}^{\rm inter} \approx \sum_{i=N}^{190} E_{i,j}^{\rm
  dimer, inter}$.\\

\noindent
For training the ML-PES for the DCM dimers,
PhysNet\cite{MM.physnet:2019}; utilizing the ML-PES fitting
environment Asparagus\cite{MM:asparagus:2025} was employed. The
dataset consisted of 4000 DCM-monomer structures and 38000 DCM-dimers
sampled from the 200 distinct clusters resulting in a total of 42000
structures. The ML-PES was trained on reference energies, forces,
dipoles and charges. All the reference calculations were carried out
at the DLPNO-MP2/cc-pVTZ level of theory using ORCA.\cite{ORCA}\\

\noindent
To smoothly combine these two regimes, a cutoff ($r_{cut} = 8$ \AA\/)
was employed. For $r_{cut} \leq 8$ \AA\/, dimer energies were those
from the ML-PES, whereas for $r_{\rm cut} \geq 8$ \AA\/ the dimer
interaction energy ($E^{\rm dimer,NB}_{i}$) was evaluated using the
CGenFF charges and LJ-parameters, see Eq. \ref{eq4}. The cutoff
($r_{cut}$) was sampled from the C-C distance in a DCM dimer at an
interval of 1 \AA\/ from 3 \AA\/ to 18 \AA\/ and the choice of cutoff
was based on the lowest RMSE against the reference cluster energy

\begin{equation}
E_{i}^{\rm dimer,inter} =
\begin{cases}
E^{\mathrm{dimer}}_{i} - \big( E^{\mathrm{monomer}}_{a,i} +
E^{\mathrm{monomer}}_{b,i} \big), & r_i < r_{\rm cut}, \\ E^{\rm
  dimer,NB}_{i} & r_i \geq r_{\rm cut}~.
\end{cases}
\label{eq4}
\end{equation}\\

\noindent
The performance of the different energy functions for DCM considered
here is reported in Figure \ref{fig:dcm-nano}. The conventional CGenFF
energy function features RMSE, MAE and maximum error of [3.75, 3.08,
  7.90] kcal/mol, respectively, with $R^2 = 0.924$ relative to
reference data from DLPNO-MP2/cc-pVTZ calculations. For an empirical
energy function this is rather good performance. However, improvements
are possible by readjusting the LJ-parameters, see Figure
\ref{fig:dcm-nano} (CGenFF + Fit) which changes the statistical
measures to [1.88, 1.56, 4.62] kcal/mol, and $R^2 = 0.936$. In other
words, all errors are reduced by a factor of $\sim 2$. Using the
ML-PES, which contains information about monomer deformation energies
and 2-body intermolecular interactions between monomers, the
performance measures are [1.57, 1.49, 2.94] kcal/mol, and $R^2 =
0.996$. Hence, from the perspective of errors between reference data
and model, the ML-PES is already better than the readjusted empirical
energy function. Most notably, the correlation coefficient is close to
1. However, the quality of the ML-PES can be further improved by
readjusting the LJ-parameters, which is shown in Figure
\ref{fig:dcm-nano} (PhysNet + Fit). This decreases the errors to
$[0.48, 0.40, 0.99]$ kcal/mol, and $R^2 = 0.997$, which is close to
chemical accuracy.\\

\noindent
The results for DCM demonstrate that ML-PESs and empirical energy
functions can be combined in a consistent manner to arrive at
high-accuracy representation of the total energy functions. Further
possibilities to boost such models is to replace, for example, the
CGenFF point charges by more elaborate representations of the
electrostatic interactions, such as atom-centered multipoles or
different flavours of distributed charge
models.\cite{MM.mtp:2013,MM.mtp2:2016,MM.dcm:2014,MM.fmdcm:2022,MM.kmdcm:2024}
Most importantly, such approaches can also be applied to larger
monomers than DCM and multicomponent molecular systems. \\

\subsection{Pure Water}
Water is essential for life and involved in much of terrestrial, and
interstellar,
chemistry.\cite{Wolfenden:2011,Lynch:2020,Gallo:2016,Clark:2010,Cisneros:2016}
As a material, water in the condensed phase is also famous for its
many anomalies: Despite its low molecular weight of 18 g/mol, water
possesses a boiling point of 372 K and reaches a maximum density 4
degrees above its freezing temperature of 277 K, while exhibiting a
high surface tension and high
viscosity.\cite{Wolfenden:2011,Lynch:2020,Gallo:2016,Clark:2010,Cisneros:2016}
These anomalies arise in part due to the hydrogen-bonding capabilities
between neighbouring water molecules in the condensed phase. For
realistic computational modeling, the relevance of anisotropic
intermolecular interactions (e.g. directionality of the H-bond)
usually requires use of models beyond simple atom-centered point
charges. Such models strive at describing higher-order multipolar
interactions which can be accomplished in different
ways.\cite{ren:2003,troester:2013,qi:2015,sidler:2018}\\

\begin{figure}
    \centering
    \includegraphics[width=0.99\linewidth]{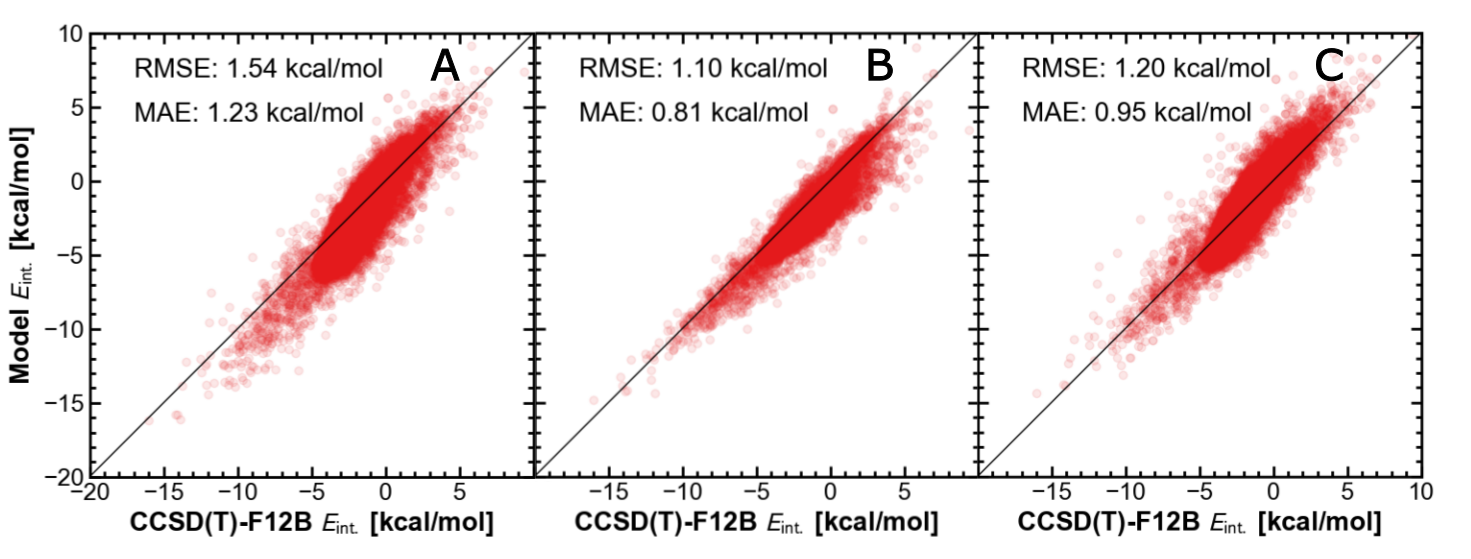}
     \caption{Performance of model energy functions versus interaction
       energies $E_{\rm int}$ calculated using the supermolecular
       approach at the CCSD(T)-F12B/dev-2zp level of theory for water
       dimers, trimers, and tetramers; using (A) TIP3P unoptimized,
       (B) MDCM with refit LJ parameters and (C) TIP3P with
       reoptimized LJ parameters.\cite{MM.water:2025}}
     \label{fig:watertetra}
\end{figure}

\noindent
Water, as a paradigmatic ``complex liquid'', is an ideal system for
developing and testing ML-based workflows.  Accurately capturing
water’s phase properties is a formidable challenge.  Research in water
modeling for MD simulations generally follows two main directions:
capturing molecular interactions with high precision by incorporating
many-body interactions (usually up to
four-body),\cite{yu:2022,zhu2023mb} or alternatively, scalable models
designed to handle large system sizes efficiently usually by fitting
the model parameters to best reproduce experimentally determined
quantities such as density, heat of vaporization and
self-diffusion.\cite{tip3p} A complete many-body description offers an
exceptionally accurate description of water but are often limited in
their application to smaller system sizes (typically 256 monomers) due
to their high computational cost. On the other hand, the mean-field,
two-body empirical force field approximation, as is usually selected,
usually suffer deficiencies in the PES that may become manifest in
inconsistent dynamics (inaccurate rotational self-correlation
life-times), and some thermodynamic properties discussed later. Within
the present work, the performance of using small and medium-sized
water clusters in developing a machine learning-based energy function
is compared for a number of candidate solutions after optimizing the
LJ-parameters, see Figure \ref{fig:watertetra}\\

\begin{figure}
    \centering
    \includegraphics[width=0.99\linewidth]{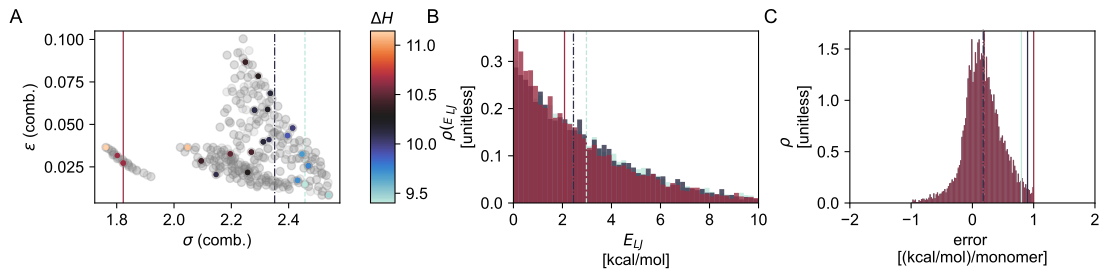}
    \caption{Panel A: The distribution of LJ parameters obtained from
      fitting to water cluster data (dimers, trimers and tetramers
      calculated at the CCSD(T)-F12B/dev-2zp level of theory). A
      selection of parameters were simulated in CHARMM and the
      resulting $\Delta H$ was obtained in units of kcal/mol. Three
      models were selected (red, black, and blue vertical lines) with
      comparable RMSEs $\sim$ 1.0 (kcal/mol)/monomer on the entire
      dataset. Panel B: Distribution of the LJ energy contributions
      for the three parameter sets. Dashed vertical lines report the
      mean of each distribution. Panel C: Error (per monomer)
      distributions for the three parameter sets. Solid vertical lines
      report the RMSE on the training distribution. The experimental
      value $\Delta H$ is 10.5 kcal/mol. As expected, the $\Delta H$
      predicted by the model is largely influenced by the magnitude of
      the average of the LJ contribution.}
    \label{fig:vdw1}
\end{figure}

\noindent
Here, a generic cluster-based workflow based on a combination of
machine learning-based and empirical representations of intra- and
intermolecular interactions was used.\cite{MM.water:2025} The total
energy is decomposed into internal contributions, and electrostatic
and van der Waals interactions between monomers. For the monomer
potential energy surface a small neural network is combined with
intermolecular interactions described by a flexible, minimally
distributed charge model and van der Waals interactions. This differs
from DCM fro which standard atom-centered CGenFF charges were used for
the electrostatics. Remaining contributions between reference energies
from electronic structure calculations and the model are fitted to
standard Lennard-Jones (12-6) terms.\\

\noindent
For water as a topical example, reference energies for the monomers
are determined from CCSD(T)-F12 calculations whereas for an ensemble
of cluster structures containing $[2,60]$ and $[2,4]$ monomers DFT and
CCSD(T) energies, respectively, were used to best match the van der
Waals contributions. Based on the bulk liquid density and heat of
vaporization, the best-performing set of LJ(12-6) parameters was
selected and a wide range of condensed phase properties were
determined and compared with experiment. Figure \ref{fig:vdw1}A
reports all fitted LJ-parameters in the $(\sigma,epsilon)-$plane and
colored points provide computed heat of formation $\Delta H$. It can,
for example, be seen that larger atom radii $(\sigma)$ yield lower
$\Delta H$ (blue) whereas decreasing $\sigma$ brings $\Delta H$ into
better agreement with the measured value of 10.5 kcal/mol. Figures
\ref{fig:vdw1}B and C report the distribution of LJ-interactions for
all parameter sets shown in Figure \ref{fig:watertetra} and the error
distributions between reference calculations and fitted model for the
three models.\\

\subsection{Eutectic Mixtures}
Deep eutectic mixtures (DEM) - also referred to as deep eutectic
solvents (DESs) when used in practical applications - are
multicomponent systems consisting of molecules acting as hydrogen bond
acceptors and hydrogen bond donors at particular molar
ratios.\cite{abbott2003DES,marcus2019trends,martins2019defdes} One of
the distinguishing features of DESs is that the melting point of the
mixture is lower than that of the individual components, due to, for
example, charge delocalization occurring through hydrogen bonding
between anions and hydrogen donors.\cite{smith:2014} Such mixtures can
also contain ions which leads to pronounced crowding and strong
electrostatic interactions, similar to ionic
liquids,\cite{Arriaga:2019} and DESs are also of interest in the
context of batteries and fuel cells due to the high cryostability,
thermal stability, and their electrochemical
stability.\cite{Sloovere:2022,Hariyanto:2023,Zhou:2024} The particular
mixture considered here consists of water, acetamide (ACEM) and NaSCN
which is present as solvated Na$^+$ and SCN$^-$ (thiocyanate)
ions. Acetamide forms low-temperature eutectics with a wide range of
inorganic salts and the resulting non-aqueous solvents have a high
ionicity.  Such mixtures have also been recognized as excellent
solvents and molten acetamide is known to dissolve inorganic and
organic
compounds.\cite{Guchhait:2010,Kalita:1998,Hu:2004,Wallace:1972} The
SCN$^-$ anion is a suitable spectroscopic probe because the CN-stretch
vibration absorbs in an otherwise empty region of the infrared
spectrum. Recently, advantage has been taken of this to probe the effect
of water addition to urea/choline chloride and in acetamide/water
mixtures.\cite{sakpal:2021,MM.eutectic:2022}\\

\begin{figure}[ht]
    \centering
    \includegraphics[width=\textwidth]{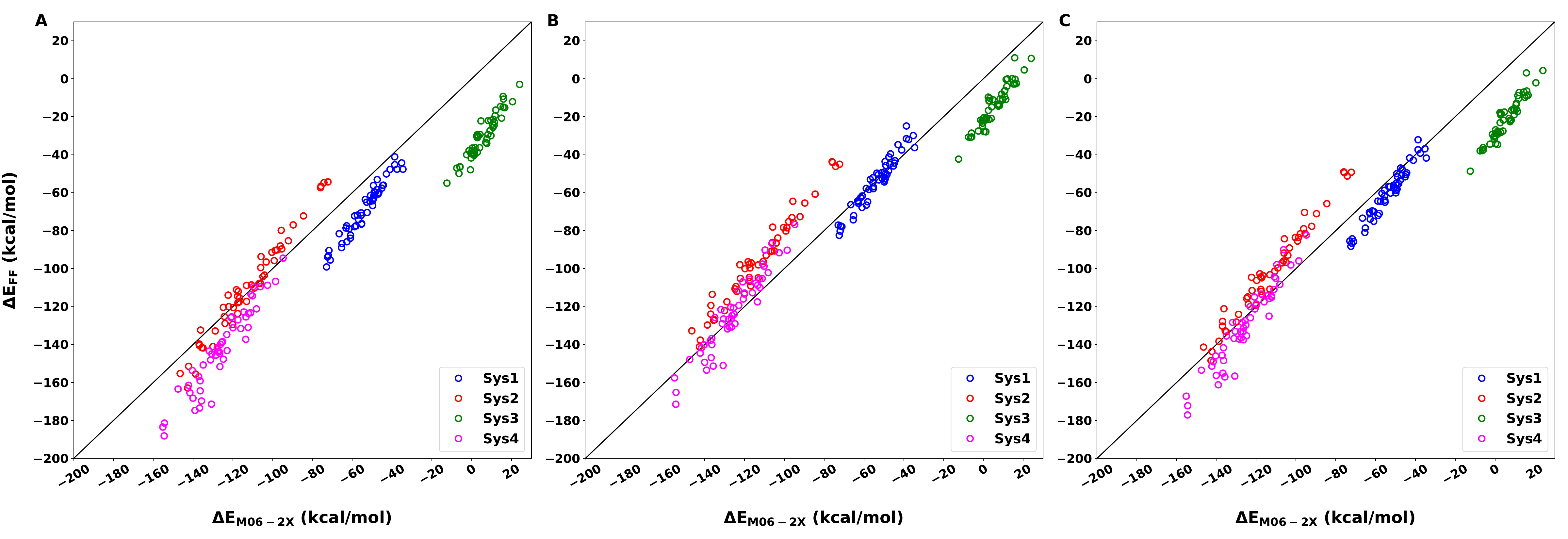}
    \caption{Correlation of interaction energies between reference DFT
      data and the empirical energy function for clusters extracted
      from simulations with [20/80] W/ACEM. Panel A: correlation
      before parameter optimization with the initial
      parameters;\cite{MM.eutectic:2022} panel B: parameters from
      individual optimization for the [20/80] mixture; panel C: using
      a transferable parameter set.}
    \label{fig:fit-all-w20}
\end{figure}

\noindent
Previously, as an initial step towards the energy functions of mixed
clusters, the dynamics of a deep eutectic mixture (KSCN/acetamide) was
studied with different W/ACEM ratios.\cite{MM.eutectic:2022} To
generate energy functions for such heterogeneous systems, a
cluster-based approach which optimized the Lennard-Jones (LJ)
parameters of SCN$^-$ to the DFT calculated energetic reference was
used.\cite{MM.des:2025} The resulting optimal parameters enable better
and more accurate predictions of viscosity and spectroscopic
properties from MD simulations. In the following a cluster-based
approach was applied to the deep eutectic mixture (NaSCN/acetamide)
with different ratio of water contents.\\

\begin{figure}[ht]
    \centering
    \includegraphics[width=0.75\linewidth]{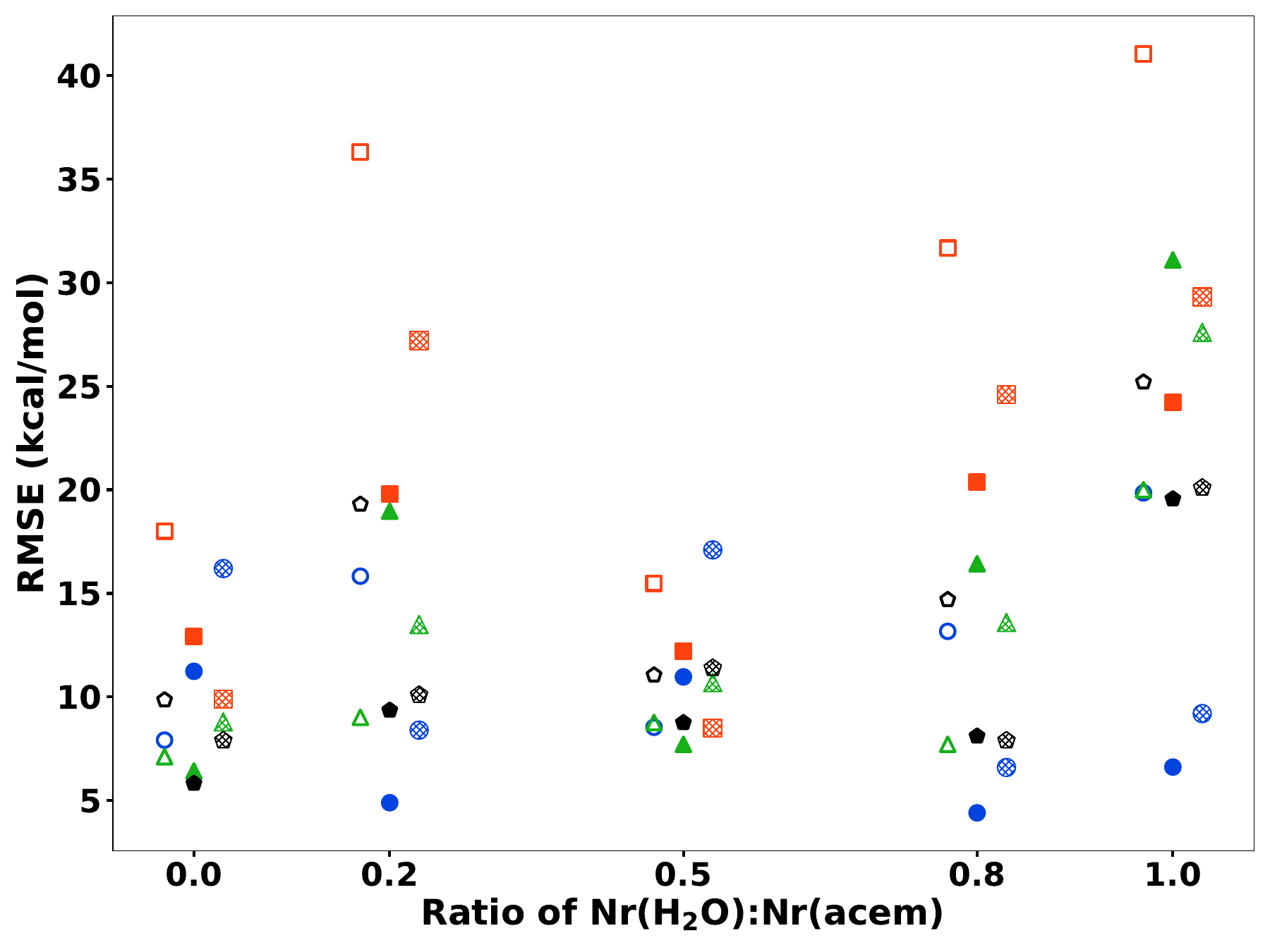}
    \caption{Summary of RMSEs between DFT calculations and final
      fitted outputs of different set of clusters from 5 mixtures with
      TIP3P water model. The hollow, filled, and hatched markers are
      for the initial, individually optimized, and transferable
      parameters. Blue, green, orange, and black symbols correspond to
      correspond to sys1 to sys4, respectively. Note that the labels
      Sys1 to Sys4 refer to specific system compositions depending on
      the W/ACEM ratio.}
    \label{fig:tip3p_rmse}
\end{figure}

\noindent
For the cluster-based optimization scheme first MD simulations of 75
Na$^+$ / 75 SCN$^-$ in 5 different water / acetamide mixtures were
carried out. The water / ACEM particle number ratios were 0.0, 0.2,
0.5, 0.8, and 1.0, see Table \ref{sitab:mixture}. In a next step, 50
clusters were extracted randomly from the MD simulations for each
mixing ratio containing one SCN$^-$ surrounded by 4 differently
organized environments (system1 to system4). The composition of these
environments (clusters) for all 5 mixtures are reported in Tables
\ref{sitab:fit-0w} to \ref{sitab:fit-100w}. For each mixture,
snapshots of the simulation were screened with respect to combinations
of [Na$^+$, SCN$^-$, ACEM, water] with a cutoff range of 5 \AA\/
around the central SCN$^-$ and 50 cluster structures for Sys1 to Sys4
were extracted.\\

\noindent
For each of the 200 clusters total interaction energies were
calculated at the M062X/AVTZ level of theory using
Gaussian16\cite{gaussian16}. In a next step, the total interaction
energy was determined from the mixed ML/MM energy function, where the
ML potential describes monomer energies and the MM potential describes
all nonbonded interactions, and the LJ-parameters $\varepsilon$ and
$r_{\rm min}$ of the three atoms SCN$^-$, were optimized using the
truncated Newton (TNC) algorithm for the 200 clusters. The final
LJ-parameters after fitting are reported in Table
\ref{tab:ljfit-para}. Two different strategies were pursued. In the
first, ``individually optimized'' LJ-parameters were determined for
each of the 5 mixtures considered. This can be regarded as the maximum
refinement level possible. However, such a parametrization scheme is
not particularly useful if one wants to investigate mixtures of
arbitrary W/ACEM combinations. For this, a more ``transferable'' set
of parameters is more useful which yields acceptable accuracy for any
amount of W and ACEM in a particular mixture. It should be further
noted that a large number of possible solutions of the minimization
problem exist for 6 free parameters (LJ-parameters for SCN$^-$) and
200 structures of various compositions.\\

\begin{table}[ht]
    \centering
    \begin{tabular}{l|c|c|c|c|c|c|c}
    \hline\hline
    \multirow{2}{*}{\makecell{LJ\\params}} & \multirow{2}{*}{Initial} &  \multicolumn{5}{c|}{Individually optimized (W/A)} & \multirow{2}{*}{Transferable} \\ 
     & & [0/100] & [20/80] & [50/50] & [80/20] & [100/0] & \\ \hline
    $\epsilon(\rm S)$ & -0.364 & -0.427 & -0.0115 & -0.568 & -0.0245 & -0.392 & -0.250 \\
     $\Tilde{r}_{\rm min}(\rm S)$ & 2.18 & 2.30 & 2.88 & 2.21 & 2.79 & 2.32 & 2.40 \\
    \hline
    $\epsilon(\rm N)$ & -0.0741 & -0.0351 & -0.0149 & -0.0955 & -0.0306 & -0.0728 & -0.0100 \\
     $\Tilde{r}_{\rm min}(\rm N)$ & 2.01 & 2.08 & 2.35 & 2.06 & 2.03 & 2.27 & 2.35 \\
     \hline
    $\epsilon(\rm C)$ & -0.102 &  -0.200 & -0.00165 & -0.183 & $-1\times10^{-4}$ & $-1\times10^{-4}$ & -0.102 \\
     $\Tilde{r}_{\rm min}(\rm C)$ & 1.79 & 1.50 & 2.08 & 1.50 & 1.94 & 1.69 & 1.79 \\
    \hline\hline
    \end{tabular}
    \caption{The LJ parameters of SCN$^-$ before and after fitting for
      the different [W/ACEM] mixtures. Two types of optimizations were
      considered. "Individually optimized" refers to optimization of
      the LJ-parameters of 200 clusters (set1 to set4) for a given
      [W/ACEM] mixture whereas "Transferable" refers to parameters
      that were initially obtained from fitting selected clusters for
      the [20/80], [50/50], and [80/20] mixtures with slight manual
      readjustments.}
    \label{tab:ljfit-para}
\end{table}

\noindent
It is first noted that for the individually optimized LJ-parameters
the parameters for the carbon atom vary widely. This can be explained
by the fact that in SCN$^-$ the central C-atom is effectively shielded
from the environment due to the larger S- and N-atom. Hence, the
interaction energies are not particularly sensitive to the
LJ-parameters of the C-atom. Consequently, for the transferable
parameter set the initial parameter values were retained. Based on the
individually optimized LJ-parameters, a consensus was sought for the
S- and N-parameters. Their performance, compared with the initial and
individually optimized parameters is summarized in Table
\ref{tab:ljfit-eutectic}. Overall, the average error of the
individually optimized LJ-parameters improves by 4 kcal/mol over the
initial parameters whereas for the transferable LJ-parameters the
improvement is 2.6 kcal/mol. This is still acceptable and all errors
are heavily influenced by the quality of the TIP3P water model (see
performance for [100/0]). Hence, it is expected that replacing this
simple water model with an improved description will considerably
boost performance, see quality for [0/100]. The performance of the
fitting is a compromise over all cluster compositions and relative
orientations. Hence, RMSE-values for different solvent compositions
can depend on the system (Sys1 to Sys4) considered, see
Figure~\ref{fig:tip3p_rmse}.\\

\begin{table}[h!]
    \centering
    \begin{tabular}{c|c|c|c|c}
    \hline\hline
    \multicolumn{2}{c|}{Mixture (W/A)} & Initial & Individual & Transferable \\
    \hline
    $[0/100]$ & & 10.7 & 9.1 & 10.7 \\
     & sys1 & 7.9 & 11.2 & 16.2 \\
     & sys2 & 7.1 & 6.4 & 8.8 \\
     & sys3 & 18.0 & 12.9 & 9.9 \\
     & sys4 & 9.9 & 5.8 & 7.9 \\
     \hline
    $[20/80]$ & & 20.1 & 13.3 & 14.8 \\
     & sys1 & 15.8 & 4.9 & 8.4 \\
     & sys2 & 9.0 & 19.0 & 13.5 \\
     & sys3 & 36.3 & 19.8 & 27.2 \\
     & sys4 & 19.3 & 9.4 & 10.1 \\
     \hline
    $[50/50]$ & & 11.0 & 9.9 & 11.9 \\
     & sys1 & 8.5 & 11.0 & 17.1 \\
     & sys2 & 8.8 & 7.7 & 10.7 \\
     & sys3 & 15.5 & 12.2 & 8.5 \\
     & sys4 & 11.1 & 8.8 & 11.4 \\
     \hline
    $[80/20]$ & & 16.8 & 12.3 & 13.2 \\
     & sys1 & 13.2 & 4.4 & 6.6 \\
     & sys2 & 7.7 & 16.4 & 13.6 \\
     & sys3 & 31.7 & 20.4 & 24.6 \\
     & sys4 & 14.7 & 8.1 & 7.9 \\
     \hline
    $[100/0]$ & & 26.5 & 20.4 & 21.6 \\
     & sys1 & 19.9 & 6.6 & 9.2 \\
     & sys2 & 20.0 & 31.1 & 27.6 \\
     & sys3 & 41.0 & 24.2 & 29.3 \\
     & sys4 & 25.2 & 19.6 & 20.1 \\
    \hline
    Overall & & 17.0  & 13.0  & 14.4 \
    \end{tabular}
    \caption{Average RMSE for Table~\ref{tab:ljfit-para}. Fitting
      individual mixtures reduces all RMSE whereas for the
      transferable parameters the RMSE is typically lower than for the
      initial parameters except for the [50/50] mixture. Note that for
      different [W/ACEM] mixtures sys1 to sys4 contain different
      numbers of W and ACEM molecules, see Tables \ref{sitab:fit-0w}
      to \ref{sitab:fit-100w}.}
    \label{tab:ljfit-eutectic}
\end{table}

\begin{figure}[ht]
    \centering \includegraphics[width=\linewidth]{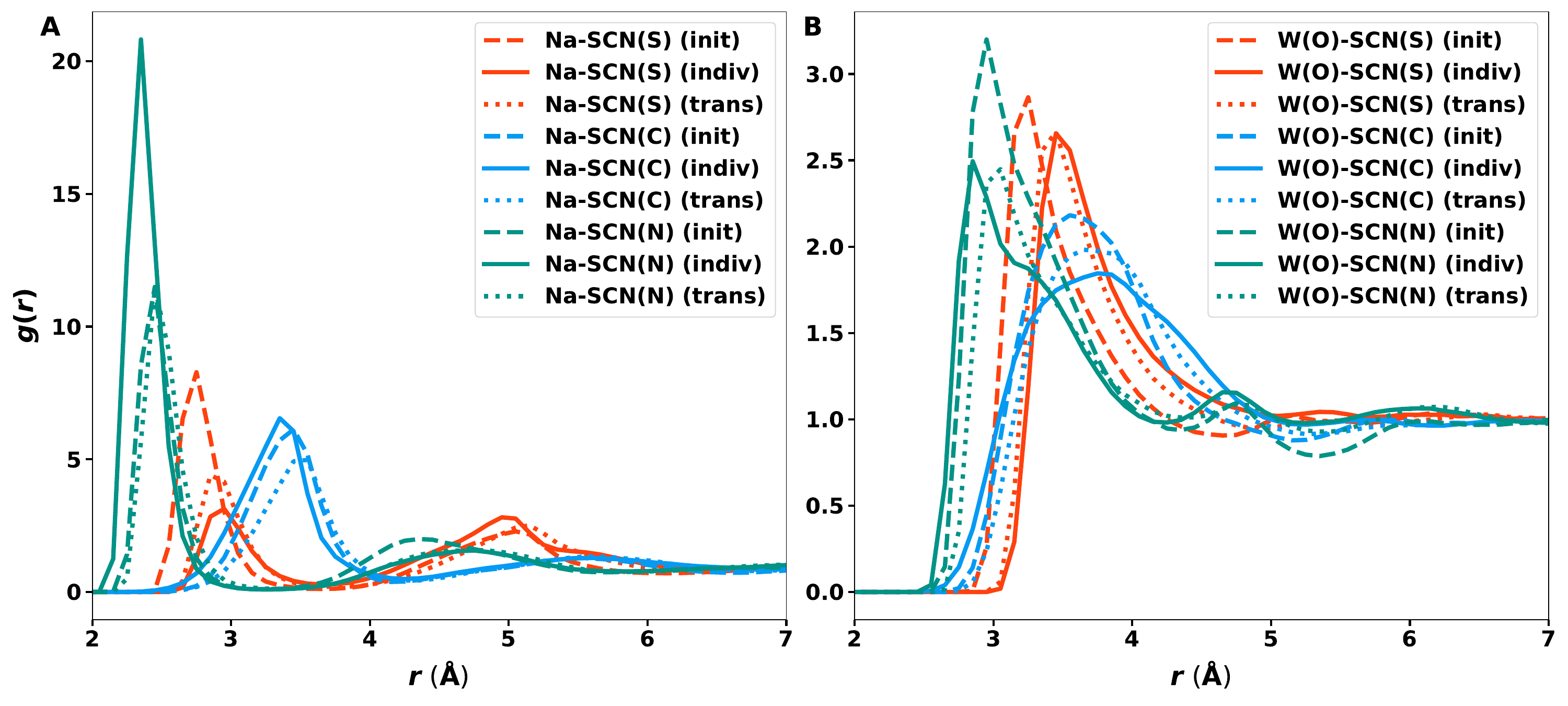}
    \caption{Comparison of the radial pair distribution function
      $g(r)$ between OW--X (X = S, C, N) with original (dashed lines)
      Lennard-Jones parameters, optimized (solid lines) Lennard-Jones
      parameters, and optimal (dotted) Lennard-Jones parameters from
      the 80/20 W/ACEM mixture. Comparison of the radial pair
      distribution function $g(r)$ between Na$^+$--X (X = S, C, N)
      with original (dashed lines) Lennard-Jones parameters, optimized
      (solid lines) Lennard-Jones parameters, and optimal (dotted)
      Lennard-Jones parameters from the 80/20 W/ACEM mixture.}
    \label{fig:gr-w80}
\end{figure}

\noindent
To quantify the effect of different parameter sets on physical
observables, changes in the Na$^+$--SCN$^-$ pair distribution
functions before and after reparametrization are reported in Figure
\ref{fig:gr-w80}. These radial distribution functions $g(r)$ were
determined from simulations of the [80/20] mixture of different
length, also to monitor convergence. Due to the large number of Na$^+$
/ SCN$^-$ pairs present, 2 ns simulations were deemed sufficient in
all cases. Dashed, solid and dotted traces in Figure \ref{fig:gr-w80}
refer to simulations using the initial, individually optimized, and
transferable LJ-parameters. The red, blue and green traces are for
separations involving the S-, C-, and N-atoms of the SCN$^-$ anion. As
Figure \ref{fig:gr-w80}A shows, the maximum peak positions for the
individually optimized parameters shorten for the Na--N$_{\rm SCN}$
and Na--C$_{\rm SCN}$ separations and increase for Na--S$_{\rm
  SCN}$. Using the transferable parameters, the $g(r)$ for the
Na--S$_{\rm SCN}$ separation is close to that for the individually
optimized set whereas $g(r)$ for Na--N$_{\rm SCN}$ differs little for
the initial data set.\\

\noindent
The peak height of $g(r)$ is a measure of the interaction strength
between the atoms involved. This indicates that for the individually
optimized parameters the Na--N$_{\rm SCN}$ is considerably stronger
than for the two other parameter sets whereas for Na--C$_{\rm SCN}$
this differs little and for Na--S$_{\rm SCN}$ individually optimized
and transferable parameters perform comparably whereas the initial
parameter set features increased interaction strength.\\

\noindent
For the interaction between water-oxygen atoms and each of the
constituent atoms of the anion, see Figure \ref{fig:gr-w80}B,
differences between parameter sets also occur but are less pronounced
overall. In general, performance of the inidividually optimized and
transferable parameter sets is comparable, whereas the initial data
set features shorter bond lengths and increased interaction
strength. This is particularly seen for the O$_{\rm W}$--N$_{\rm SCN}$
separation.\\

\noindent
As required, individually optimized LJ-parameters for the SCN$^-$
probe molecule for each mixing ratio outperform the initial and
transferable parametrizations. Nevertheless, the transferable data set
provides a meaningful description of the intermolecular
interactions. It should be noted that merely minimizing the loss
function does not necessarily yield parameters within expected ranges,
e.g. $r_{\rm min}/2$ for the sulfur atom increases up to 2.88 \AA\/
which is unusually large compared with standard values in
CGenFF.\cite{cgenff:2010} Such effects are due to both, the
mathematical description of the van der Waals interactions and the
type and composition of the reference data set. Further improvement of
the models will require a better description of the water-water
interactions, and possibly a different mathematical description of the
van der Waals interactions, while transferability may be improved by adding
higher order terms such as polarization that respond to changes in highly
polar chemical environments such as eutectic mixtures.\\

\subsection{CO on Amorphous Solid Water}
Surface processes are primary for the genesis of molecules in the
interstellar medium. In cold molecular clouds, dust behaves as a
suitable substrate for the deposition and chemical synthesis of
molecules in a bottom-up manner. The dominant form of interstellar ice
is amorphous solid water (ASW),\cite{wakelam:2017, hagen:1981} whose
intrinsically disordered hydrogen-bond network gives rise to a broad
variety of adsorption sites and binding
environments.\cite{bovolenta:2025co} This structural complexity makes
ASW a challenging substrate to model, yet it also places it at the
center of astrochemistry, as it governs how molecules adhere, diffuse,
and react on grain surfaces.\cite{cuppen:2017} Among the species of
astrochemical interest, carbon monoxide (CO) is of particular
interest, serving both as the main carbon reservoir\cite{tielens:2013}
and as the second most abundant molecule in molecular clouds.\\

\noindent
To capture these microscopic processes, simulations must describe the
interaction between adsorbed molecules and the ice surface with very
high accuracy. This requirement is especially stringent at the low
temperatures of molecular clouds ($10-50$ K), where even small errors in
adsorption energies can lead to large deviations in desorption rates,
diffusion barriers, and ultimately, reaction
kinetics.\cite{cuppen:2017} Moreover, proper modeling must also
account for interaction within the ice H$_2$O molecules: during
molecular formation, excess reaction energy couples back into the
water network for energy dissipation, altering its local structure and
influencing subsequent reactivity. Thus, an accurate description must
encompass both intermolecular interactions between adsorbates and ASW,
as well as intramolecular interactions within the ice
matrix.\cite{hama:2013surface}\\

\begin{figure}[ht]
    \centering
    \includegraphics[width=0.99\linewidth]{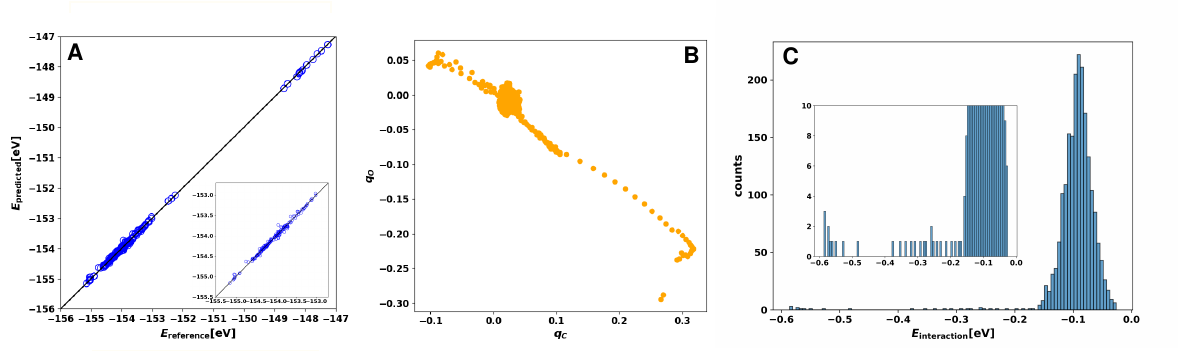}
    \caption{Panel A: Correlation between predicted and reference
      energies for a test set of 195 CO/water clusters, with an RMSE
      of 0.0399 eV and an MAE of 0.0279 eV. The low-energy structures
      are from MD simulations at 50 K using the KKY water
      model,\cite{mm:2024co2, kky_orig} while the high-energy
      structures were generated using xTB\cite{bannwarth2019gfn2} to
      add higher-energy configurations. Panel B: Atomic charges of the
      C and O atoms in CO across all 2007 clusters on ASW calculated
      using NN model. Panel C: Distribution of interaction energies
      for all 2007 clusters at M062X/aug-cc-pVTZ + D3 level.}
    \label{fig:co+asw_corr}
\end{figure}
  
\noindent
Traditionally, MD simulations have employed hybrid approaches
combining QM (or ML) with MM with either mechanical or electrostatic
embedding.\cite{gao:1996methods,senn:2009,MM.oxy:2019,mm:2024co2}
While these hybrid approaches reduce computational cost, they rely on
empirical potentials that sacrifice accuracy. This presents challenges
in accurately capturing the diverse, heterogeneous adsorption
environments on the ASW and the energy redistribution effects (i.e.,
coupling between internal degrees of freedom of the adsorbate and the
adsorbent) essential for understanding surface chemistry. In contrast,
fully data-driven pure ML potentials trained directly on \textit{ab
  initio} data offer a more accurate and internally consistent
framework which is done for CO on ASW. In other words, the total
interaction energy of the clusters in question is represented using a
NN-PES.\\

\noindent
To determine the appropriate cluster size, the interaction energy
($E_{\rm int}$) between CO and water clusters containing 9 to 18 water
molecules was computed at the M06-2X/aug-cc-pVTZ +
D3\cite{zhao2008m06,grimme:2010} level using ORCA.\cite{ORCA,ORCA5}
The results in Figure \ref{sifig:Eintvswatermol} show that $E_{\rm
  int}$ varies noticeably with cluster size due to the influence of
cavity shape and local water arrangement. Balancing accuracy and
computational cost, 14 water molecules were selected for dataset
generation. Cluster structures were extracted from long MD simulations
of CO diffusion on the water surface (from earlier
work\cite{mm:2024co2}) to capture the diverse arrangements and
cavities present on ASW. Energies, forces, and dipole moments were
calculated for 2007 clusters, and the dataset was trained using
Asparagus\cite{MM:asparagus:2025} with an 80/10/10 split for training,
validation, and test sets. The correlation between the predicted and
reference energies for the test set is shown in Figure
\ref{fig:co+asw_corr}A. Panel \ref{fig:co+asw_corr}B shows the atomic
charges of the C and O atoms in CO, and panel \ref{fig:co+asw_corr}C
the interaction energy range with amorphous solid water, both
evaluated across 2007 clusters. Strong interactions occur mainly in
cavities, which explains the low probability of highly negative
interaction energies in panel \ref{fig:co+asw_corr}C.

\subsection{Menshutkin Reaction}
The Menshutkin reaction is a key $S_N2$ reaction in organic and
bioorganic chemistry and has been widely studied since the original paper
back in 1890.\cite{menschutkin:1890,MM.mensh:2022} In this reaction,
neutral reactants form ionic products carrying opposite charges. It is
known that the reaction behaves differently in solution compared to
the gas phase; it proceeds much faster in polar solvents than in less
polar ones. \cite{su:2007} While it is clear that solvent effects
strongly influence the reaction energetics, much less is known about
the molecular details of the process, particularly about the solvent
dynamics and reorganization along the reaction
path. \cite{MM.mensh:2022,dutta:2020,aziz:2024}\\

\noindent
In this part, the reaction between NH$_3$ and CH$_3$Cl was
investigated in the presence of $n = 0, 1, 2,$ and $5$ explicit water
molecules using a machine-learning-based potential energy surface
(ML-PES). Similar to CO on ASW the total interaction energy was
represented using PhysNet. To build the training dataset, three types of structures were
generated: (i) normal-mode samples, (ii) metadynamics snapshots, and
(iii) local minima obtained via the GOAT algorithm.\cite{de:2025}\\

\begin{figure}
    \centering
    \includegraphics[width=1\linewidth]{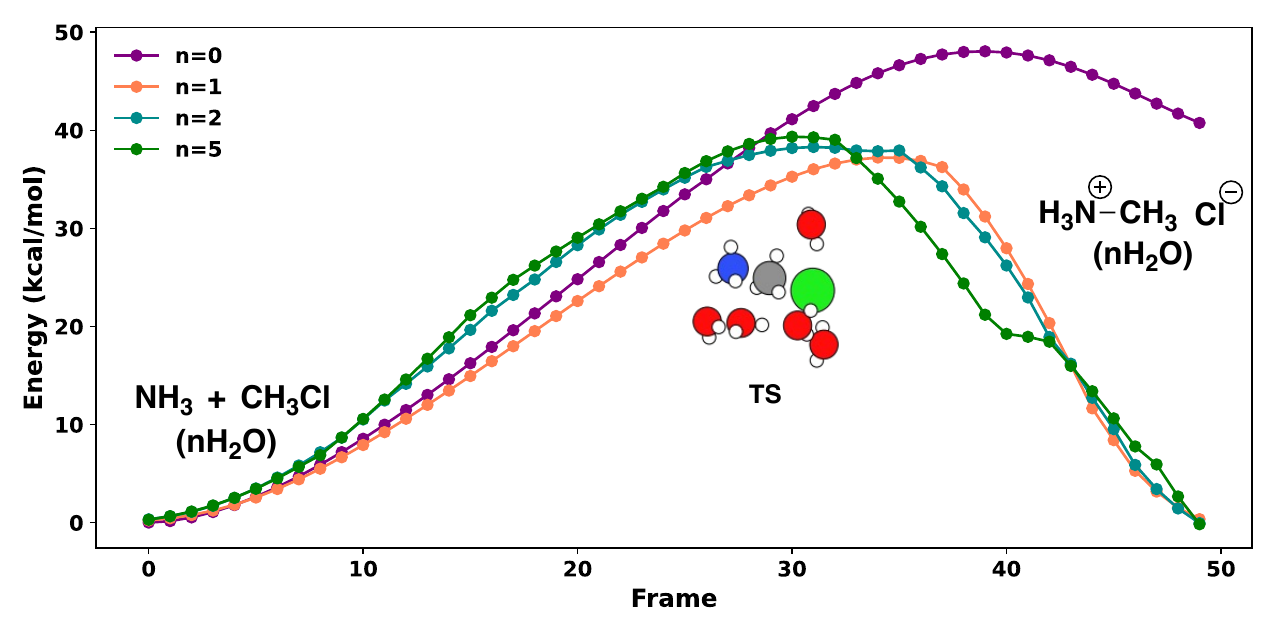}
    \caption{One-dimensional energy
      profile along the N-C and C-Cl bond for Menshutkin reaction with
      different number of water molecules. Predominantly, the first half of the reaction (frames 1 to 25) involves C--Cl bond elongation whereas C--N bond formation occurs during the second half. However, the two coordinates are coupled throughout the chemical transformation as is known for S$_N$2 reactions. The energy function is the
      ML-PES represented using PhysNet.}
    \label{fig:menshenergy}
\end{figure}

\noindent
The reaction paths for systems with varying numbers of water molecules
were first determined using the NEB-TS algorithm, as implemented in
the ORCA software package, at the B3LYP/ma-def2-TZVP level of
theory.\cite{ORCA,becke:1993,rappoport:2010} All generated structures
were then used for normal-mode sampling at 100, 300, 500, and 1000 K
at the PBE/def2-SVP level of theory using the Asparagus
software.\cite{MM:asparagus:2025} In addition, metadynamics
simulations were carried out with GFN2-xTB across the same temperature
range. \cite{bannwarth:2019}\\

\noindent
Because solvent organization plays a key role in understanding
reaction mechanisms and the role of solvation, the GOAT algorithm was
further applied to identify local minima corresponding to different
cluster organizations. Altogether, this procedure yielded 33,300
structures, which were subsequently used to determine energies, forces
and dipole moments at the RI-MP2/cc-pVTZ+cc-pVTZ/C level of theory
using ORCA.\cite{weigend:1998} The resulting dataset was then employed
to train a machine-learning model within the Asparagus framework.\\

\begin{figure}[h!]
    \centering
    \includegraphics[width=1\linewidth]{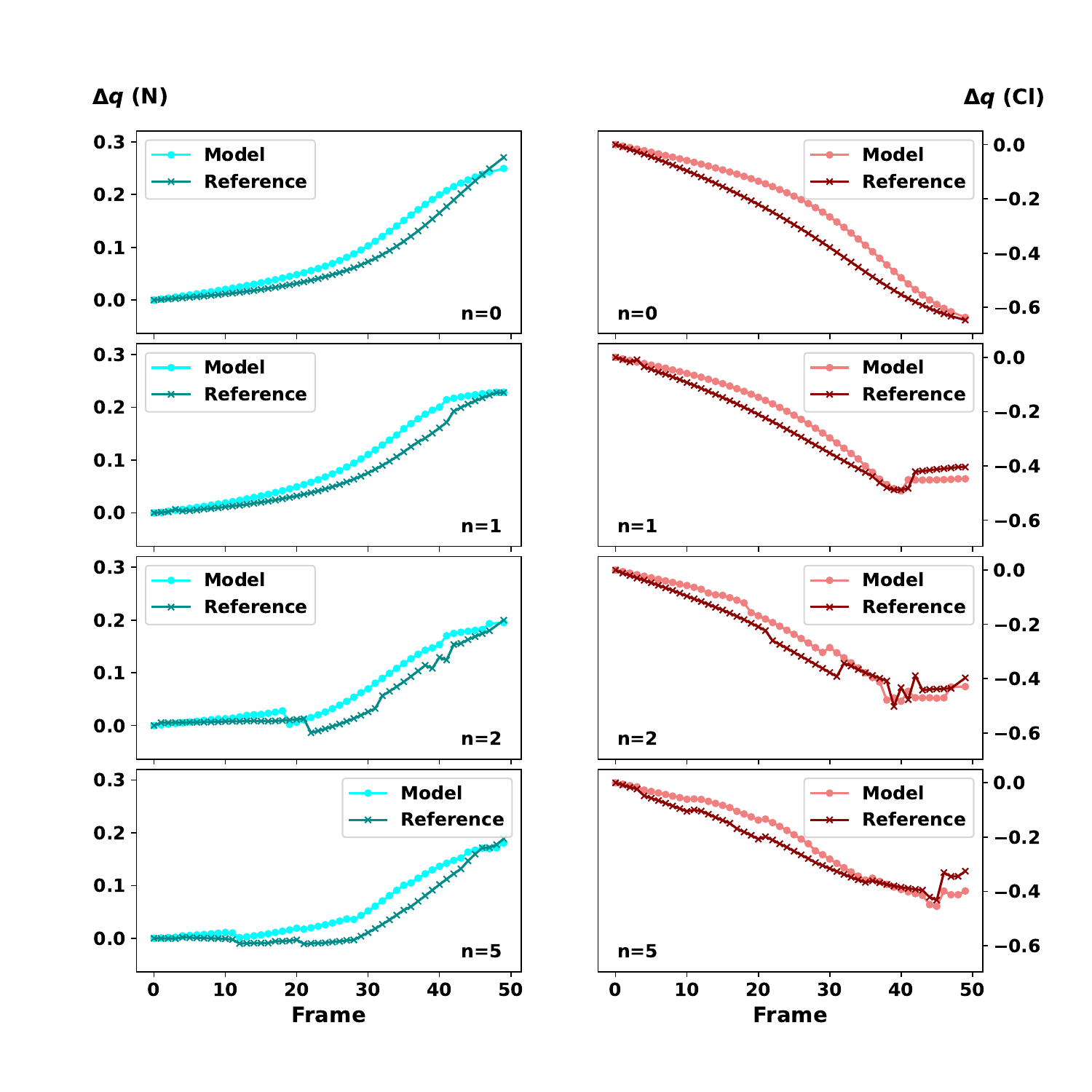}
    \caption{Comparison between model predicted and Hirshfeld charge
      evolution on the N atom (left panel) and on the Cl atom (right
      panel) along one-dimensional scan for the Menshutkin reaction
      with different number of water molecules.}
    \label{fig:menshcharges}
\end{figure}

\noindent
These steps yielded a stable ML model capable of simulating systems
with varying numbers of surrounding water molecules. The values of MAE
are $0.22$ kcal/mol for energies and $0.35$ kcal/mol$\cdot$\AA\/ for
forces, while corresponding RMSE values are $0.74$ kcal/mol and $1.23$
kcal/mol$\cdot$\AA\/, respectively, indicating the high quality of the
trained model. Using this model, a one-dimensional scan was carried
out along the reaction coordinate defined by the N–C and C–Cl
separations. The scan started at an N–-C distance of 3.2 \AA\/ and a
C–-Cl distance of 1.8 \AA\/, and the trajectory was divided into 50
frames up to final distances of 1.48 \AA\/ (N–C) and 3.0 \AA\/
(C–Cl). During this process, the N, C, and Cl atoms were fixed, while
the remainder of the system was allowed to relax. All optimizations
were performed using the ML-PES with the ASE
framework.\cite{larsen:2017}\\

\noindent
The paths including one or several water molecules provide an
impression for the effect of solvation for a gas-phase–like reaction,
see Figure~\ref{fig:menshenergy}. The results clearly show that in the
absence of coordinating solvent water the charged product is
significantly destabilized. However, addition of a single water
molecule already increases the overall stability of the product by
$\sim 10$ kcal/mol. Notably it is found that the barrier height does
not strongly depend on the number of water molecules included. It
should be noted, however, that a 1-dimensional picture is does not
provide a complete description of this reaction.\\

\noindent
Because PhysNet provides (fluctuating) atomic charges it is also of
interest to monitor changes on the N- and Cl-atoms along the
1-dimensional reaction, providing insight into charge redistribution
along this reaction path, see Figure~\ref{fig:menshcharges}. Setting
the reactant state as the reference, close agreement between the
charge evolution along the reaction pathway, as predicted by the ML
model, and the Hirshfeld charges \cite{spackman:2009} calculated for
each snapshot at the RI-MP2/cc-pVTZ+cc-pVTZ/C level of theory is
found. The results also show that fluctuating charges are required to
realistically describe such a reaction both, in the gas phase and in
solution. The change in the partial charges on the N- and Cl-atoms
depends on the degree of hydration. Again, the first water molecule
has the largest influence. It reduces the charge on the N- and
Cl-atoms between reactant and product state by 30 \%. Addition of one
or 4 water molecules does not change the charge in the product state
but influences the curvature of the curve on the reactant side. The
observed decrease in the N- and Cl-charges can be attributed to the
presence of water molecules in the outer coordination sphere of the
reactants and products. As illustrated in the figure, this effect
becomes more apparent with an increasing number of water
molecules. These molecules act as a compensating network, thus
facilitating charge redistribution by providing additional
contributors to the overall charge distribution. For the nitrogen
atom, the charge remains nearly constant up to the transition state. A
noticeable increase in charge is observed only after the transition
state, indicating the role of the solvent in facilitating charge
compensation. This is consistent with earlier work on the Menshutkin
reaction.\cite{MM.mensh:2022}\\

\subsection{Spectroscopic Probes}
Spectroscopic probes are small molecules that can be used to label
proteins or ligands for characterization of the energetics and
dynamics in condensed phase environments. Specific examples include
cyanide (--CN), azide (--N$_3$), or nitric oxide (--NO). Importantly,
these small molecules exhibit infrared signatures that set them
clearly apart from the vibrational modes of the guest molecule. As an
example, protein vibrational spectra extend up to $\sim 1800$
cm$^{-1}$, followed by a largely "empty" region up to 2800 cm$^{-1}$
above which the X-H stretch vibrations are located (X = C, N,
O). Hence, by modifying residues such as alanine using such a label,
its spectroscopy and vibrational dynamics can be followed with great
precision. For example, IR spectroscopy can distinguish stretching
frequencies associated with CN$^-$ or
N$_3^-$,\cite{fafarman:2006,meuwly-scn:2024,meuwly-n3:2019} which
reveal differences in bond order and electron delocalization. By
employing these spectroscopic probes, the characteristic properties of
these species can be unraveled to understand their roles in various
fields of chemistry.\\

\noindent
In this last example the behavior of -SCN, -SNO, and -N$_3$ probes in
different water cluster environments was investigated. To establish a
consistent framework, methyl-substituted reference systems (Me-X) were
constructed, with CH$_3$ group Lennard-Jones (LJ) parameters obtained
from the CHARMM\cite{MM.charmm:2024} force field via
CGenFF\cite{cgenff:2010}, while the probe atoms were explicitly
parapmetrized. For each system, a dedicated fMDCM charge
model\cite{MM.fmdcm:2022} was generated. The total charge of each
system was set to zero. The resulting electrostatics were employed to
refine the Lennard-Jones (LJ) parameters, thereby ensuring a balanced
description of nonbonded interactions.\\

\begin{figure}
    \centering
    \includegraphics[width=1\linewidth]{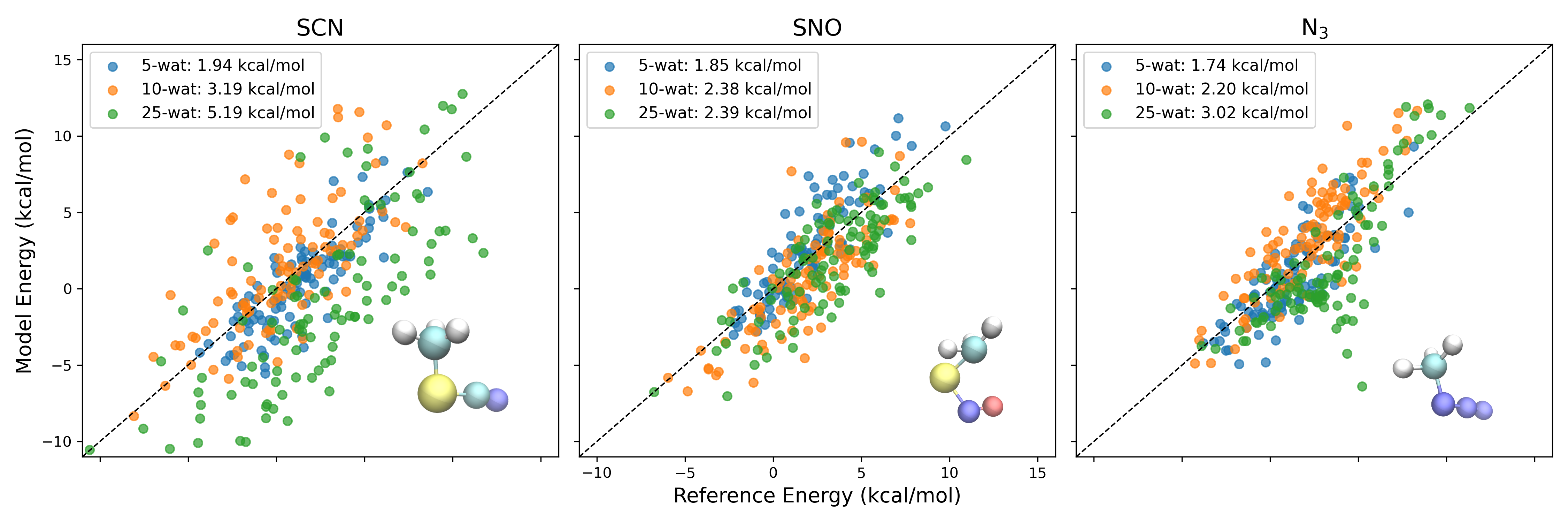}
    \caption{Correlation between reference interaction energies from
      B3LYP/aug-cc-pVDZ calculations with the model interaction
      energies for CH$_3$-X clusters using the fMDCM electrostatic
      model. Blue, orange, and green colors represent the clusters
      that contain 5, 10, and 25 waters, respectively.}
    \label{fig:specprobe}
\end{figure}

\noindent
To account for solvation effects, probe-water clusters of varying
sizes, containing 5, 10, and 25 water molecules, were examined. From
molecular dynamics simulations, 100 configurations were randomly
sampled for each cluster, with the exception of the CH$_3$-SCN 10- and
25-water cluster (99 configurations), the CH$_3$-SNO with 5- and
10-water clusters (98 and 97 configurations, respectively), and the
CH$_3$-N$_3$ 25-water cluster (96 configurations). Total interaction
energies for all configurations were subsequently computed using
Gaussian\cite{gaussian16} at the B3LYP/aug-cc-pVDZ level of theory.
This was the reference data for fitting the LJ parameters within the
CHARMM framework in conjunction with the fMDCM model for each
spectroscopic probe. For the fitting, curve fit in SciPy
\cite{2020SciPy-NMeth} was used, which minimizes the sum of squared
residuals between the model and the reference data using the
non-linear least-squares solver. CHARMM nonbonded interaction energies
were compared to quantum chemical cluster energies with monomer
energies removed, similar to DCM and water nonbonded energies
described above.\

\noindent
The fitted LJ parameters, $\epsilon$ and $r_{min}/2$ for each
water-cluster-specific probe are summarized in Tables
\ref{tab:probe_lj_scn} to \ref{tab:probe_lj_n3}. The tables also
report the atomic charges obtained from the fMDCM model (up to 4
charges per atom). The correlation between the reference interaction
energies and those from the fitted models are shown in Figure
\ref{fig:specprobe}. In all cases the RMSE decreases considerably
compared with the initial parameters. For -SNO, -N$_3$, and -SCN
surrounded by 5 water molecules the RMSE ranges from 4.3 to 5.5
kcal/mol which decreases to 1.8 to 1.9 kcal/mol after readjusting the
LJ-parameters. Increasing the size of the solvent shell to 10 and 25
water molecules, the errors using the initial LJ-parameters increase
to $\sim 8$ kcal/mol and $\sim 13$ kcal/mol, see Table
\ref{sitab:unfitted-LJ-probes}. Hence, the error scales with the
number of water molecules which is indicative of a considerable gain
that can be expected from using improved water models in the
future. After fitting the LJ-parameters, the RMSE-values range from
2.2 to 3.2 kcal/mol (10 water molecules) and 2.4 to 5.2 kcal/mol (25
water molecules), respectively. Hence, the LJ-fitted models improve by
a factor of up to $\sim 5$ in terms of reproducing the reference
calculations.\\

\section{Conclusion}
This manuscript uses pure and mixed molecular clusters for improving
partial or full ML-based energy functions. For this, finite-sized
clusters are extracted from condensed-phase simulations. In a next
step, reference data at the highest affordable levels of quantum
chemistry are determined from which interaction energies are
obtained. This data set constitutes the reference data to optimize in
particular Lennard-Jones parameters. This is a meaningful approach
because for internal degrees of freedom (bonds, valence angles,
dihedrals - in the language of empirical energy functions) highly
accurate ML-PESs can be obtained from either kernel- or NN-based
approaches. Both have been used in the present work. For the
electrostatics on the other hand, a range of methods to best describe
the electrostatic potential are available. Here, the minimal
distributed charge models either without or with conformational
adjustments are employed. From the perspective of a non-polarizable
empirical energy function the only remaining contribution is then the
van der Waals term.\\

\begin{figure}
    \centering
    \includegraphics[width=0.85\linewidth]{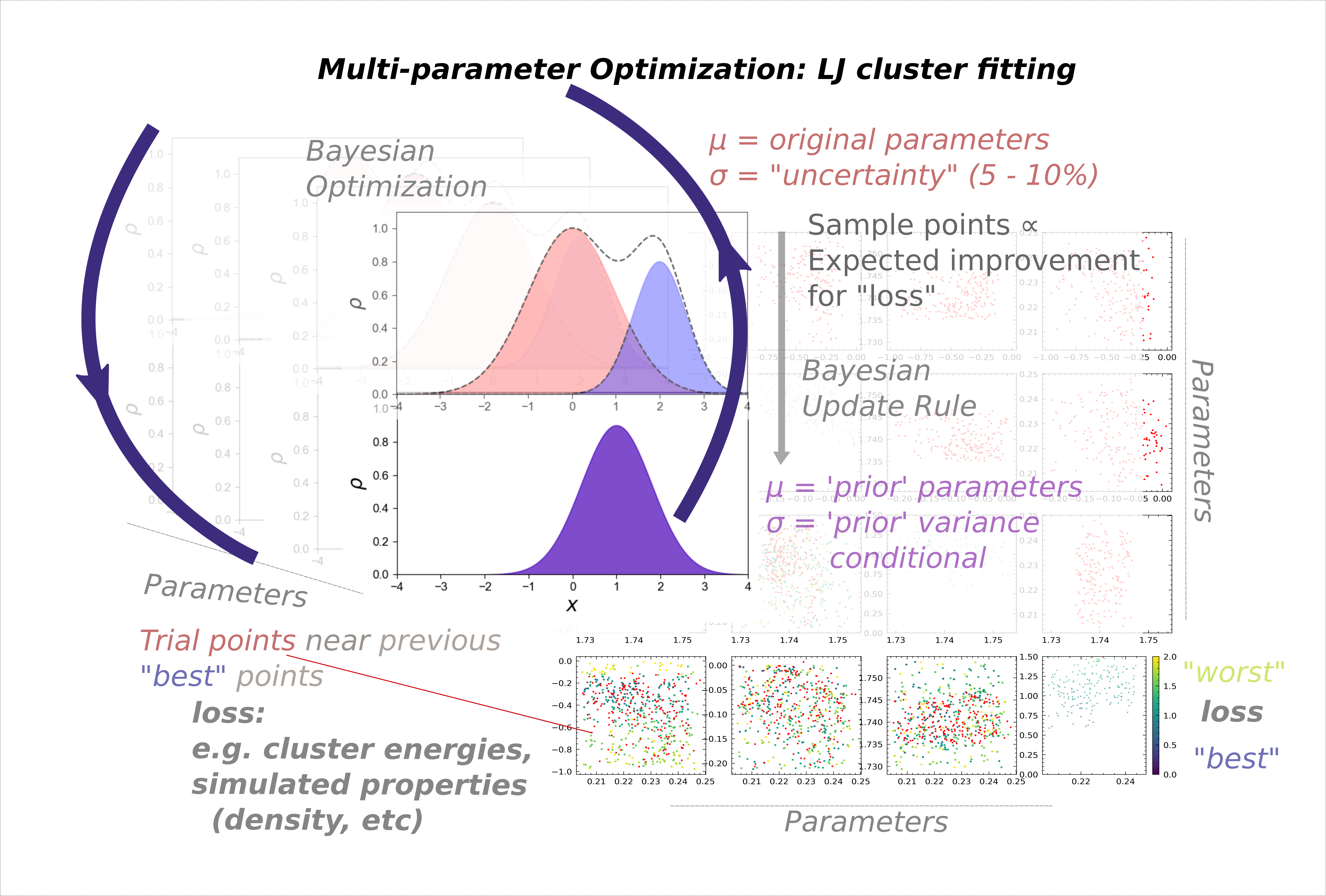}
    \caption{An uncertainty-aware, multi-parametrization optimization
      strategy for obtaining optimized LJ-parameters using cluster
      formation energies and simulated properties. Bayesian
      optimization marginalizes previous knowledge or 'priors' with
      new information, for example from MD simulations, to make
      informed decisions on new areas of parameter space to
      explore. Parameter searches can prioritize exploration (sample
      points based on variance) or exploitation (sampling based on
      expected ) depending on the updating rule for selecting trial
      points.}
    \label{fig:bayes}
\end{figure}

\noindent
Within the broader perspective to generate next-generation energy
functions for condensed phase simulations, the role of experiments
also needs to be discussed. The present work clearly shows that
representing reference data from electronic structure calculations and
(re)adjusting certain key contributions to the total energy (here the
LJ-parameters were improved) can provide qualitatively and
quantitatively improved energy functions. On the other hand, given the
large amount of available data that can be generated from {\it ab
  initio} calculations invariably leads to overdetermined fitting
problems with a multitude of competitive solutions. Constraining this
target space of equally likely solutions can be accomplished through
calculation of {\it experimental observables} and comparing with
measurements. This was, for example, done recently for
water.\cite{MM.water:2025} Earlier efforts, based on more empirical
expressions for the total energy, yielded high-quality models for
water.\cite{Wang:2013} Similarly, for the infrared spectroscopy of
trialanine in water, a Bayesian reweighting approach based on the
measured IR spectrum yielded an improved conformational ensemble in
dihedral angle space (Ramachandran plot).\cite{feng:2018}
Interesingly, this ensemble was consistent with subsequent simulations
using improved empirical energy functions which also correctly
described the IR spectrum.\cite{MM.ala3:2021} Yet an alternative
approach is to {\it morph} entire PESs to improve agreement between
computed and measured observables which has, however, only been done
for gas phase systems so far.\cite{MM.morph:1999,MM.morph:2024} There
have also been efforts to go beyond Bayesian reweighting to improve
empirical energy functions for specific systems.\cite{kofinger:2021}\\

\noindent
As has been demonstrated here, there are clear limitations due to the
functional form of the PES, e.g. parametric dependence of van der
Waals interactions, that impacts the performance of total energy
functions derived from cluster data. other challenges such as
overfitting are equally important from a technical
stand-point.\cite{wang:2014,wang:2025} Statistical approaches such as
boot-strapping are well known in the community. Related Bayesian
interpretations of convergence, e.g. of the loss function, are often
extremely helpful in this high dimensional multi-objective
optimizations; in fact any least squares optimization can be restated
as imposing a prior distribution of parameter probabilities
(i.e. Bayesian) (Figure \ref{fig:bayes}).  The prior widths,
e.g. restraining LJs parameters to be within a certain percentage of
literature values, reflect the expected variations of the parameters
during the optimization, and, in an empirical Bayes approach, can be
obtained through resampling (and updating the prior beliefs),
implemented in such efforts as Force
Balance.\cite{wang:2014,wang:2025} A Bayesian approach helps combine
cluster based energy models with different modalities of training data
(such as some desired simulated properties).\\

\noindent
As the examples discussed in the present work indicate, a
cluster-based approach yields models for improved total energies which
eventually can be used in molecular simulation (here demonstrated for
eutectic liquids). The work highlights that improvements in either
parametrized expressions for describing van der Waals interaction or
resorting to ML-based approaches for this contribution will further
boost models for intermolecular interactions in ``simple'' and
``multi-component'' molecular systems, in particular when viewed in
the context and together with experimentally measured properties
amenable to molecular simulation.\\

\section*{Supplementary Material}

\section*{Data Availability}
The codes and data for the present study are available from
\url{https://github.com/MMunibas/cluster} upon publication.

\section*{Acknowledgment}
The authors gratefully acknowledge financial support from the Swiss
National Science Foundation through grants $200020\_219779$ (MM),
$200021\_215088$ (MM), and the University of Basel (MM). This article
is also based upon work within COST Action COSY CA21101, supported by
COST (European Cooperation in Science and Technology) (to MM).\\

\clearpage
\bibliography{refs}
\clearpage

\renewcommand{\thetable}{S\arabic{table}}
\renewcommand{\thefigure}{S\arabic{figure}}
\renewcommand{\thesection}{S\arabic{section}}
\renewcommand{\d}{\text{d}}
\setcounter{figure}{0}  
\setcounter{section}{0}  
\setcounter{table}{0}

\newpage

\noindent
{\bf SUPPORTING INFORMATION: Cluster Models for Next-Generation Energy
  Functions for Molecular Simulations}

\section{Additional Figures for Water}
\begin{figure}
    \centering
    \includegraphics[width=0.75\linewidth]{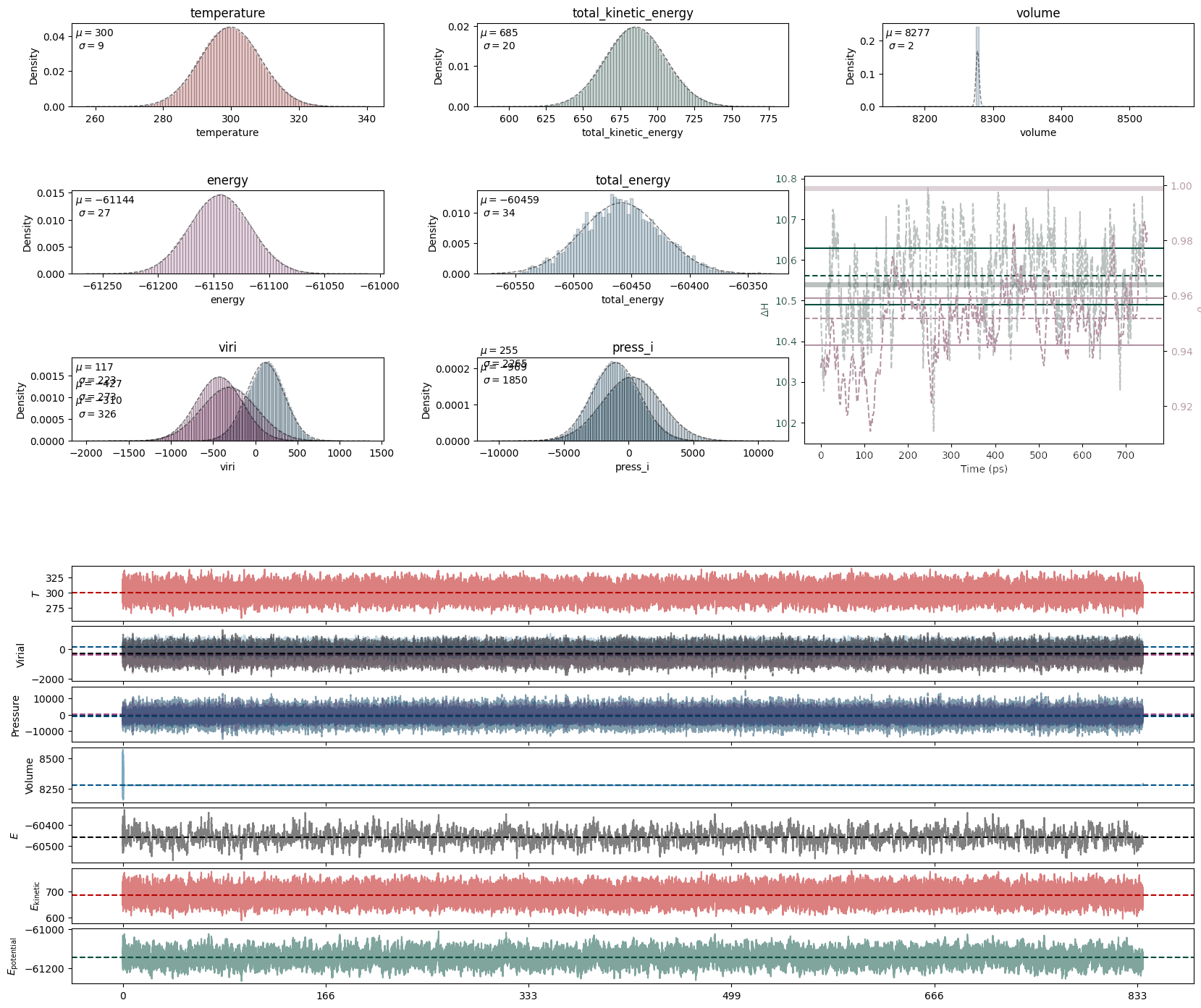}
    \caption{Refinement of LJ parameters may require a large number of
      trial simulations. For liquid water at ambient conditions, the
      predicted enthalpy of vaporization and density fluctuate with a
      few percent of the experimental values after a relatively short
      (1 ns) heating and equilibration producing stable NVT behavior
      (time series and distributions), arriving at canonical
      distributions within an additional 1 ns of simulation time. The
      predictive uncertainty associated with fluctuations before and
      after equilibration can be used to inform future trial
      parameters.  }
    \label{sifig:simunc}
\end{figure}

\subsection{Specification of eutectic mixtures}
The different mixtures studied here are specified in Table
\ref{sitab:mixture}.
\begin{table}[h]
    \centering
    \begin{tabular}{l|c|c|c}
    \hline\hline
     Index & Cation & Water percentage & Acetamide percentage \\
     \hline
     sod0 & \ce{Na+} & 0\% & 100\% \\
     sod20 & \ce{Na+} & 20\% & 80\% \\
     sod50 & \ce{Na+} & 50\% & 50\% \\
     sod70 & \ce{Na+} & 70\% & 30\% \\
     sod80 & \ce{Na+} & 80\% & 20\% \\
     sod90 & \ce{Na+} & 90\% & 10\% \\
     sod100 & \ce{Na+} & 100\% & 0\% \\
     hyb0 & 45\ce{Na+}, 30\ce{K+} & 0\% & 100\% \\
     hyb1 & 45\ce{Na+}, 30\ce{K+} & 100\% & 0\% \\
     \hline
     \hline
    \end{tabular}
    \caption{Specification of mixtures studied in the simulation}
    \label{sitab:mixture}
\end{table}

\section{LJ parameter fitting for NaSCN in TIP3P/acetamide}

\begin{table}[ht]
    \centering
    \begin{tabular}{l|c|c|c|c}
      & SCN$^-$ & Acetamide & TIP3P & Na$^+$ \\
    \hline
    sys1 & 1 & 6 & 0 & 2 \\
    sys2 & 1 & 5 & 0 & 1 \\
    sys3 & 2 & 5 & 0 & 2 \\
    sys4 & 2 & 4 & 0 & 3 \\ \hline
    \end{tabular}
    \caption{Composition of clusters with 0\% water.}
    \label{sitab:fit-0w}
\end{table}

\begin{table}[h!]
    \centering
    \begin{tabular}{l|c|c|c|c}
      & SCN$^-$ & Acetamide & TIP3P & Na$^+$ \\
    \hline
    sys1 & 1 & 5 & 1 & 0 \\
    sys2 & 1 & 4 & 2 & 1 \\
    sys3 & 2 & 5 & 1 & 0 \\
    sys4 & 2 & 4 & 2 & 2 \\ \hline
    \end{tabular}
    \caption{Composition of clusters with 20\% water + 20\% ACEM.}
    \label{sitab:fit-20w}
\end{table}

\begin{table}[ht]
    \centering
    \begin{tabular}{l|c|c|c|c}
      & SCN$^-$ & Acetamide & TIP3P & Na$^+$ \\
    \hline
    sys1 & 1 & 6 & 1 & 2 \\
    sys2 & 1 & 5 & 2 & 1 \\
    sys3 & 2 & 4 & 1 & 2 \\
    sys4 & 2 & 3 & 2 & 3 \\ \hline
    \end{tabular}
    \caption{Composition of clusters with 50\% water + 50\% ACEM.}
    \label{sitab:fit-50w-50ac}
\end{table}

\begin{table}[h!]
    \centering
    \begin{tabular}{l|c|c|c|c}
      & SCN$^-$ & Acetamide & TIP3P & Na$^+$ \\
    \hline
    sys1 & 1 & 2 & 4 & 0 \\
    sys2 & 1 & 1 & 5 & 1 \\
    sys3 & 2 & 2 & 3 & 0 \\
    sys4 & 2 & 1 & 4 & 2 \\
    \hline
    \end{tabular}
    \caption{Composition of clusters with 80\% water + 20\% ACEM.}
    \label{sitab:fit-80w}
\end{table}

\begin{table}[h!]
    \centering
    \begin{tabular}{l|c|c|c|c}
      & SCN$^-$ & Acetamide & TIP3P & Na$^+$ \\
    \hline
    sys1 & 1 & 0 & 16 & 0 \\
    sys2 & 1 & 0 & 14 & 1 \\
    sys3 & 2 & 0 & 14 & 0 \\
    sys4 & 2 & 0 & 12 & 1 \\
    \hline
    \end{tabular}
    \caption{Composition of clusters with 100\% water.}
    \label{sitab:fit-100w}
\end{table}

\begin{figure}[ht]
    \centering \includegraphics[width=\textwidth]{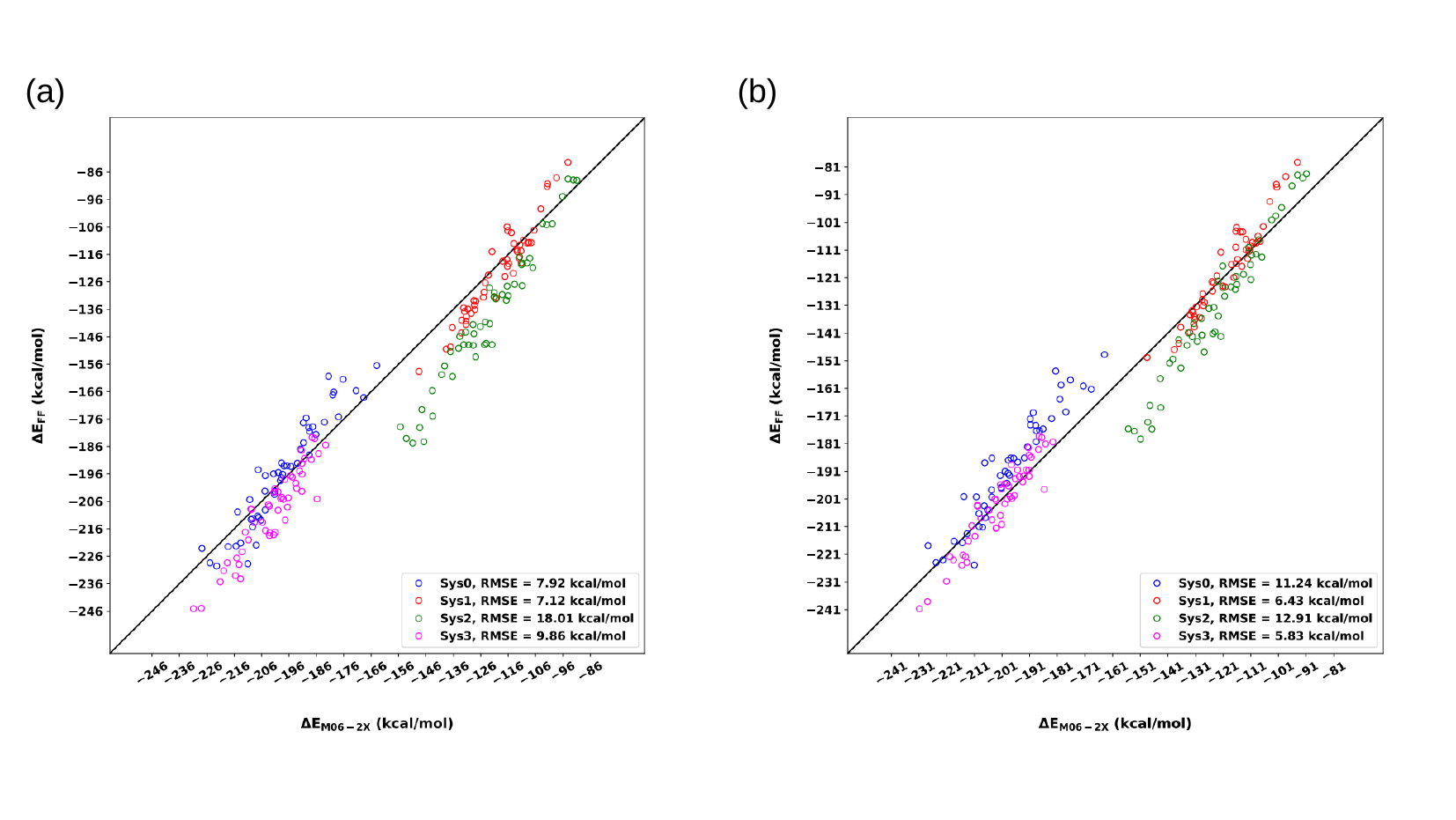}
    \caption{Correlation of interaction energies for clusters in 100\%
      acetamide solutions between calculations with (a) initial
      parameters; (b) fitted parameters and DFT.}
    \label{sifig:ljfit-eutectic}
\end{figure}

\begin{figure}
    \centering
    \includegraphics[width=0.65\linewidth]{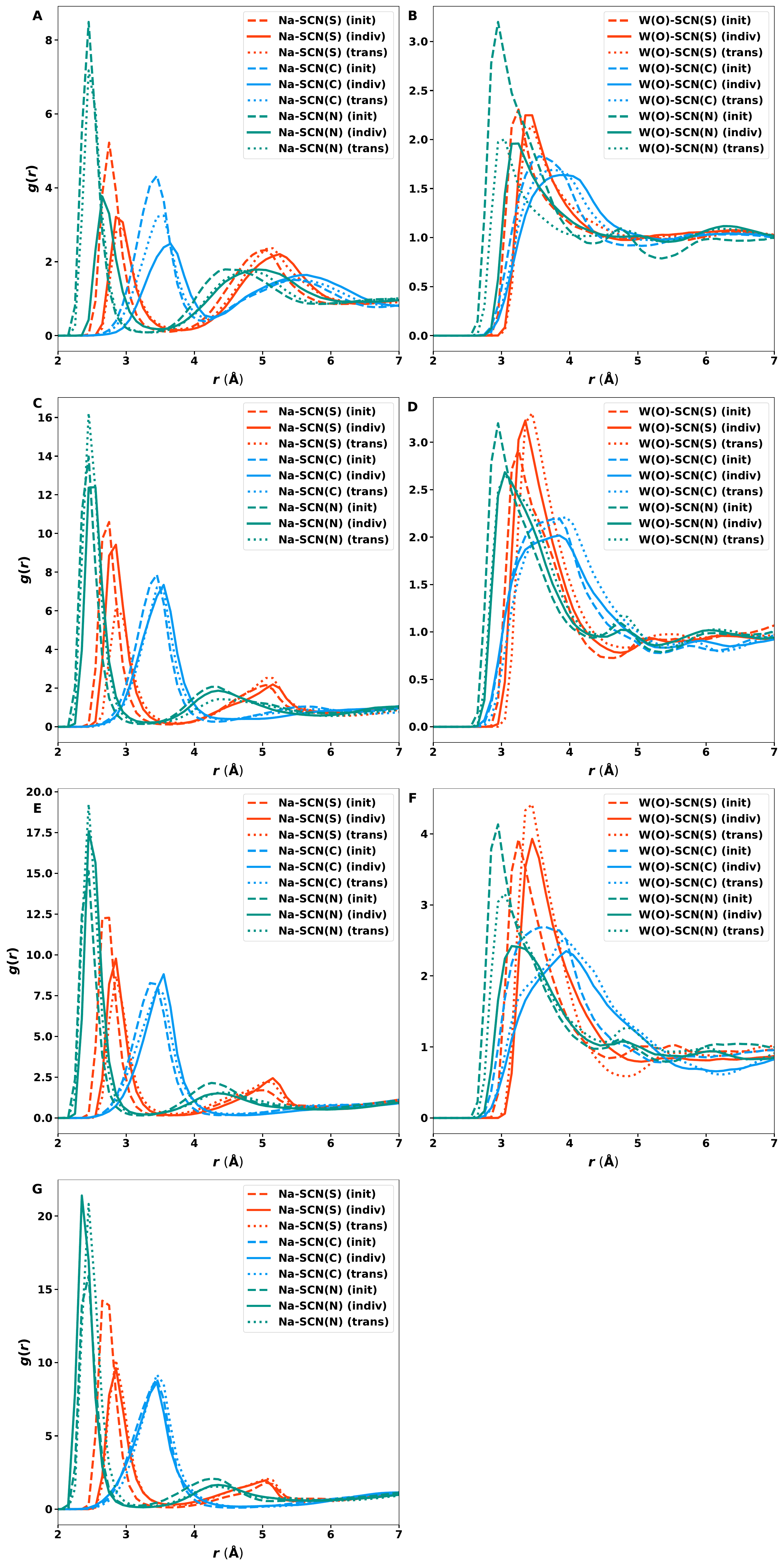}
    \caption{Comparison of the radial pair distribution function
      $g(r)$ between OW--X (X = S, C, N) with original (dashed lines)
      Lennard-Jones parameters, optimized (solid lines) Lennard-Jones
      parameters, and optimal (dotted) Lennard-Jones parameters from
      the 100/0 (A and B), 50/50 (C and D), 20/80 (E and F), and 0/100
      (G, no water) W/ACEM mixtures.}
    \label{fig:gr_rem }
\end{figure}

\begin{figure}[ht]
    \centering
    \includegraphics[width=\textwidth]{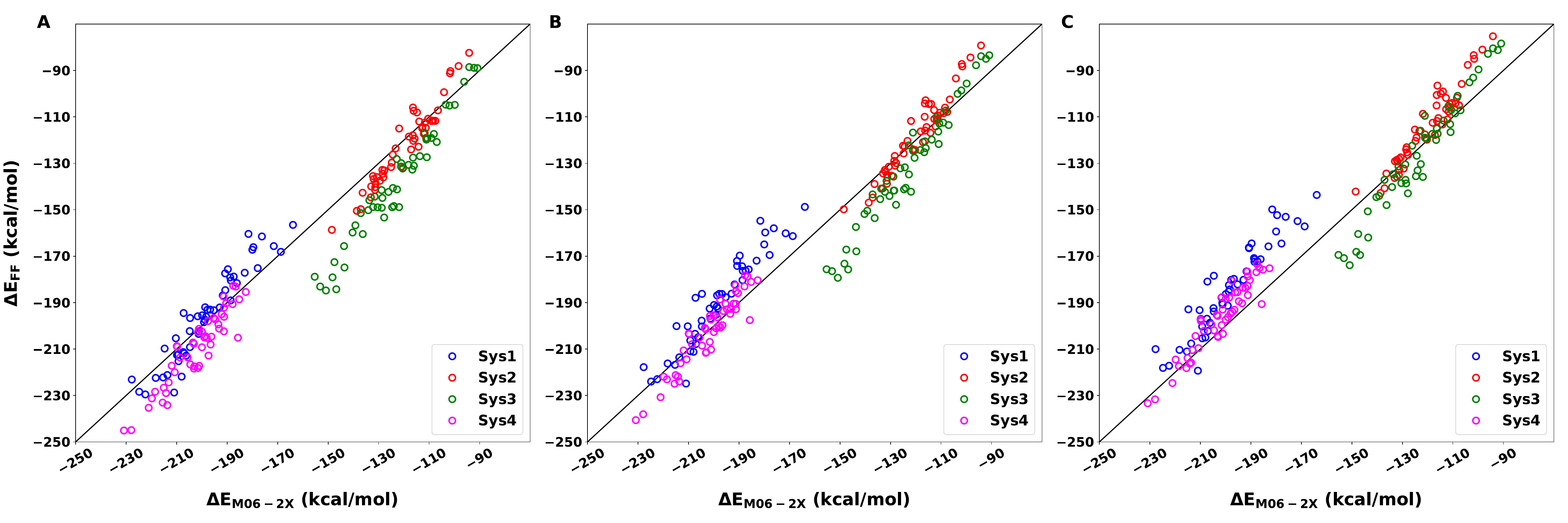}
    \caption{Correlation of interaction energies between reference DFT
      data and the empirical energy function for clusters extracted
      from simulations with [0/100] W/ACEM. Panel A: correlation
      before parameter optimization with the initial
      parameters;\cite{MM.eutectic:2022} panel B: parameters from
      individual optimization; panel C: transferable set of
      parameters.}
    \label{fig:fit-all-w0}
\end{figure}

\begin{figure}[ht]
    \centering
    \includegraphics[width=\textwidth]{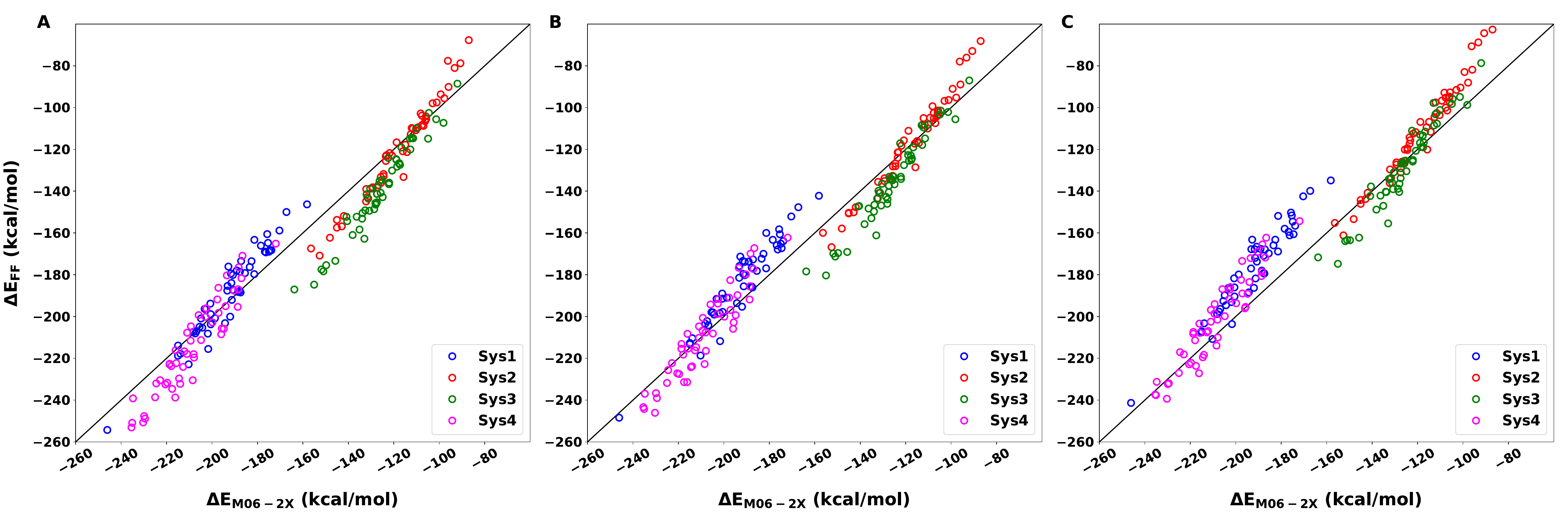}
    \caption{Correlation of interaction energies between reference DFT
      data and the empirical energy function for clusters extracted
      from simulations with [50/50] W/ACEM. Panel A: correlation
      before parameter optimization with the initial parameters; panel
      B: parameters from individual optimization; panel C:
      transferable set of parameters.}
    \label{fig:fit-all-w50}
\end{figure}

\begin{figure}[ht]
    \centering
    \includegraphics[width=\textwidth]{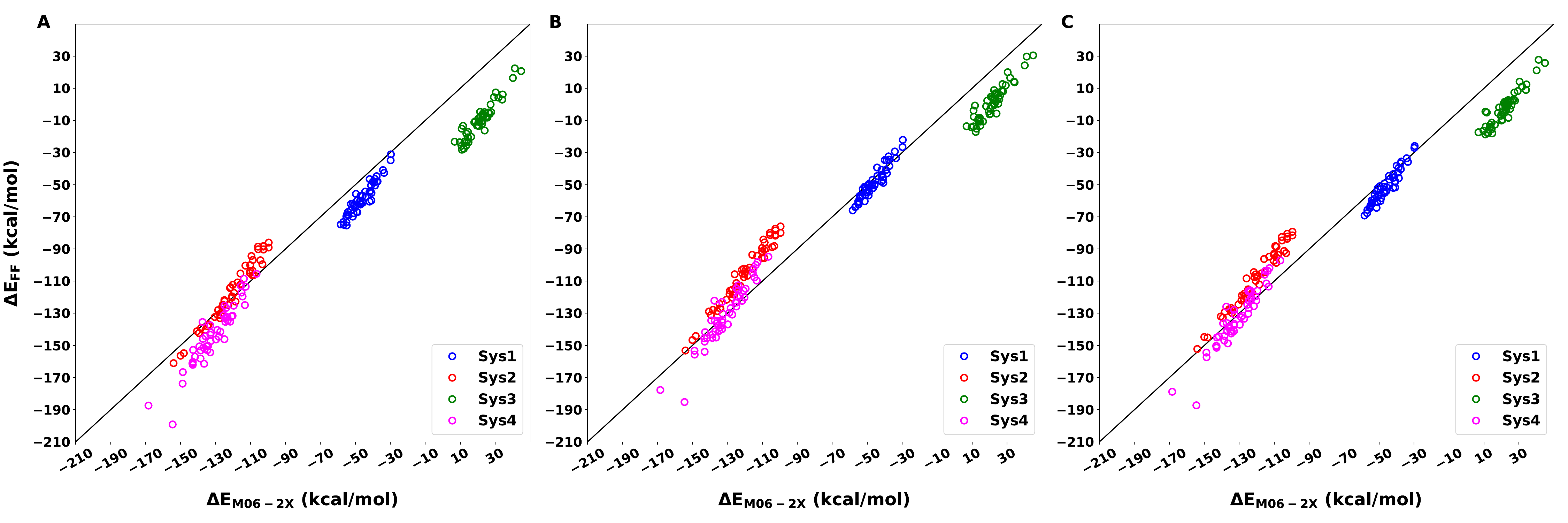}
    \caption{Correlation of interaction energies between reference DFT
      data and the empirical energy function for clusters extracted
      from simulations with [80/20] W/ACEM. Panel A: correlation
      before parameter optimization with the initial parameters; panel
      B: parameters from individual optimization; panel C:
      transferable set of parameters.}
    \label{fig:fit-all-w80}
\end{figure}

\begin{figure}[ht]
    \centering
    \includegraphics[width=\textwidth]{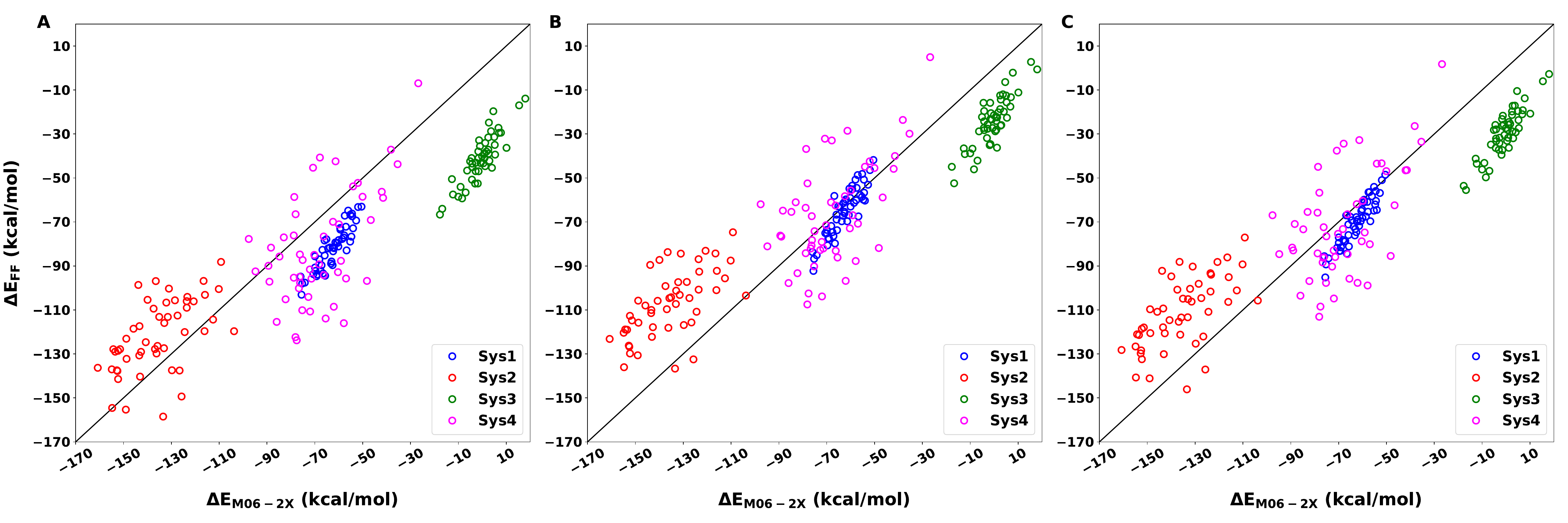}
    \caption{Correlation of interaction energies between reference DFT
      data and the empirical energy function for clusters extracted
      from simulations with [100/0] W/ACEM. Panel A: correlation
      before parameter optimization with the initial parameters; panel
      B: parameters from individual optimization; panel C:
      transferable set of parameters.}
    \label{fig:fit-all-w100}
\end{figure}

\clearpage

\section{Spectroscopic Probes}

\begin{table}[H]
    \centering
    \caption{Atomic charges from the fMDCM fit for CH$_3$SCN, along
      with Lennard-Jones (LJ) parameters obtained using the model. For
      the methyl group, LJ parameters were taken from the CGenFF force
      field.}
    \label{tab:probe_lj_scn}
    \begin{tabular}{l|c|c|c|c|c|c|c}
        \hline\hline
        \multirow{2}{*}{Atom type} 
            & \multicolumn{4}{c|}{Charges} 
            & \multicolumn{3}{c}{LJ parameters} \\ 
        & Q1 & Q2 & Q3 & Q4 & 5-wat & 10-wat & 25-wat \\ 
        \hline
        \multirow{2}{*}{S} 
        & 0.997 & -0.890 & -0.890 & 0.488 & -0.20072 & -0.6 & -0.6 \\ 
        &  &  &  &  & 1.82607 & 2.4 & 1.5 \\ 
        \hline
        \multirow{2}{*}{C} 
        & -0.305 & 0.844 &  &  & -0.18 & -0.008 & -0.008 \\ 
        &  &  &  &  & 2.17363 & 1.6 & 1.6 \\ 
        \hline
        \multirow{2}{*}{N} 
            & -0.439 & -0.864 & 0.864 & & -0.4 & -0.01 & -0.08421 \\ 
        &  &  &  &  & 1.79027 & 2.24215 & 2.4 \\
        \hline
        \multirow{2}{*}{C2} 
            & 0.010  & -0.010 &  &  &  &  &  \\ 
        &  &  &  &  &  &  &  \\
        \hline
        \multirow{2}{*}{H1-H2-H3} 
           & 0.064 &  &  & &  &  &  \\ 
        &  &  &  &  &  &  &  \\  
        \hline
    \end{tabular}
\end{table}

\begin{table}[H]
    \centering
    \caption{Atomic charges from the fMDCM fit for CH$_3$SNO, along
      with Lennard-Jones (LJ) parameters obtained using the model. For
      the methyl group, LJ parameters were taken from the CGenFF force
      field.}
    \label{tab:probe_lj_sno}
    \begin{tabular}{l|c|c|c|c|c|c|c}
        \hline\hline
        \multirow{2}{*}{Atom type} 
            & \multicolumn{4}{c|}{Charges} 
            & \multicolumn{3}{c}{LJ parameters} \\ 
        & Q1 & Q2 & Q3 & Q4 & 5-wat & 10-wat & 25-wat \\ 
        \hline
        \multirow{2}{*}{S} 
        & 1.000 & -0.636 & -0.636 & 0.070 & -0.6 & -0.6 & -0.15 \\ 
        &  &  &  &  & 1.92449 & 2.2 & 2.2 \\ 
        \hline
        \multirow{2}{*}{N} 
        & 0.463 & 0.963 & -0.236  &  & -0.23760 & -0.01 & -0.08006 \\ 
        &  &  &  &  & 2.2 & 1.0 & 2.2 \\ 
        \hline
        \multirow{2}{*}{O} 
            & -0.991 & -0.072 & -0.320 & & -0.05214 & -0.17867 & -0.01991 \\ 
        &  &  &  &  & 2.0 & 2.0 & 2.0 \\
        \hline
        \multirow{2}{*}{C2} 
            & -0.761  & 0.692 &  &  &  &  &  \\ 
        &  &  &  &  &  &  &  \\
        \hline
        \multirow{2}{*}{H1-H2-H3} 
           & 0.059 &  &  & &  &  &  \\ 
        &  &  &  &  &  &  &  \\        
        \hline
    \end{tabular}
\end{table}

\begin{table}[H]
    \centering
    \caption{Atomic charges from the fMDCM fit for CH$_3$N$_3$, along
      with Lennard-Jones (LJ) parameters obtained using the model. For
      the methyl group, LJ parameters were taken from the CGenFF force
      field.}
    \label{tab:probe_lj_n3}
    \begin{tabular}{l|c|c|c|c|c|c|c}
        \hline\hline
        \multirow{2}{*}{Atom type} 
            & \multicolumn{4}{c|}{Charges} 
            & \multicolumn{3}{c}{LJ parameters} \\ 
        & Q1 & Q2 & Q3 & Q4 & 5-wat & 10-wat & 25-wat \\ 
        \hline
        \multirow{2}{*}{N1} 
        & -0.931 &  0.448 & -1.000 &  & -0.05 & -0.05 & -0.04352 \\ 
        &  &  &  &  & 2.26148 & 2.21190 & 2.4 \\ 
        \hline
        \multirow{2}{*}{N2} 
        & 0.992 & 0.796 &  &  & -0.01 & -0.01 & -0.03102 \\ 
        &  &  &  &  & 2.32741 & 2.32741 & 2.4 \\ 
        \hline
        \multirow{2}{*}{N3} 
            & -0.887 & -0.658 & 0.910 & & -0.01 & -0.01 & -0.01 \\ 
        &  &  &  &  & 2.21731 & 2.19236 & 2.19236 \\
        \hline
        \multirow{2}{*}{C2} 
            & 0.505  & -0.301 &  &  &  &  &  \\ 
        &  &  &  &  &  &  &  \\
        \hline
        \multirow{2}{*}{H1-H2-H3} 
           & 0.042 &  &  & &  &  &  \\ 
        &  &  &  &  &  &  &  \\        
        \hline
    \end{tabular}
\end{table}

\begin{table}[h!]
    \centering
    \begin{tabular}{l|c|c|c}
      & CH$_3$SCN & CH$_3$SNO & CH$_3$N$_3$ \\
    \hline
    5-wat & 4.32 & 5.12 & 5.49  \\
    10-wat & 7.73 & 8.53 & 8.10  \\
    25-wat & 13.08 & 14.13 & 11.59  \\
    \hline
    \end{tabular}
    \caption{RMSE values (kcal/mol) of the fMDCM electrostatic models
      with unfitted LJ-parameters from CGenFF, relative to reference
      QM interaction energies. Columns show different molecules, rows
      are different cluster sizes.}
    \label{sitab:unfitted-LJ-probes}
\end{table}

\section{CO on Amorphous Solid Water}
\begin{figure}[H]
    \centering
    \includegraphics[width=0.9\linewidth]{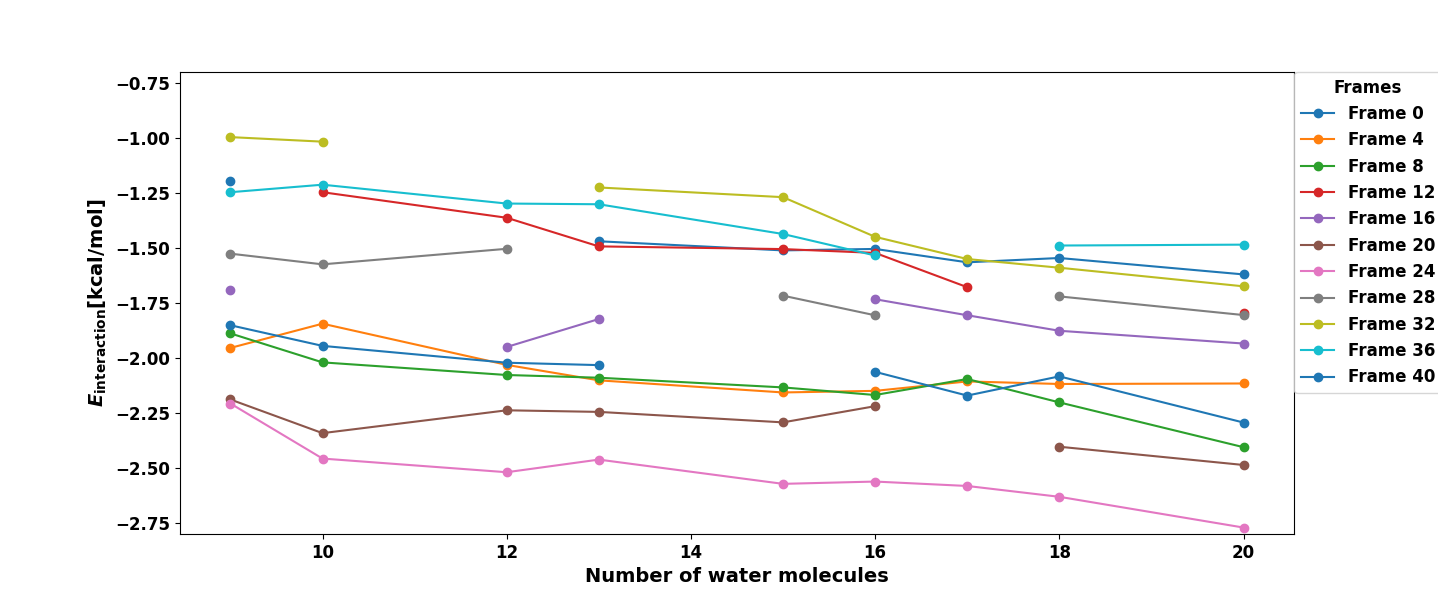}
    \caption{$E_{\rm int}$ v/s number of water molecules at
      M06-2X/aug-cc-pVTZ + D3 level for 11 different frames
      (orientations from MD simulations).}
    \label{sifig:Eintvswatermol}
\end{figure}

\end{document}